# Meteorites and the physico-chemical conditions in the early solar nebula

Jérôme Aléon[1]

**Abstract.** Chondritic meteorites constitute the most ancient rock record available in the laboratory to study the formation of the solar system and its planets. Detailed investigations of their mineralogy, petrography, chemistry and isotopic composition and comparison with other primitive solar system samples such as cometary dust particles have allowed through the years to decipher the conditions of formation of their individual components thought to have once been free-floating pieces of dust and rocks in the early solar nebula. When put in the context of astrophysical models of young stellar objects, chondritic meteorites and cometary dust bring essential insights on the astrophysical conditions prevailing in the very first stages of the solar system. Several exemples are shown in this chapter, which include (1) high temperature processes and the formation of chondrules and refractory inclusions, (2) oxygen isotopes and their bearing on photochemistry and large scale geochemical reservoirs in the nebula, (3) organosynthesis and cold cloud chemistry recorded by organic matter and hydrogen isotopes, (4) irradiation of solids by flares from the young Sun and finally (5) large scale transport and mixing of material evidenced in chondritic interplanetary dust particles and samples returned from comet Wild2 by the Stardust mission.

## 1. Introduction

Meteorites are extraterrestrial rocks that reach the Earth after ejection from their original parent-body. Based on the spectral characteristics of meteorite families and asteroids and on the calculated trajectories of several well-documented falls, most meteorites are thought to come from the asteroidal belt, although for some specific meteorites (e.g. the Orgueil meteorite, Gounelle et al. 2006a) a cometary origin has been repeatedly proposed based on various arguments. Most meteorites are as old as the solar system and underwent many events related to the formation of solids in the solar system, to their subsequent evolution into planets, and to their late processing due to parent-body geology, regolith history, interplanetary transit after ejection from the parent-body, atmospheric entry and residence on Earth. To study the formation of the solar system, it is necessary to disentangle the first nebular step from the following using laboratory analyses, whose interpretations are not always unequivocal. There are thus many sources of confusion and only a little number of samples are valuable to study the physical conditions in the protosolar nebula.

[1]Centre de Spectrométrie Nucléaire et de Spectrométrie de Masse, Orsay, France ; Jerome.Aleon@csnsm.in2p3.fr

Meteorites are usually classified as differentiated and undifferentiated meteorites. Differentiated meteorites are fragments of parent-bodies large enough to have undergone planetary differentiation. These meteorites are thus samples of core, mantle or crust of early planetesimals and are extremely valuable to understand the early formation and differentiation stages of rocky planets. Undifferentiated meteorites escaped planetary formation and are more useful to study the early solar nebula. They are commonly termed chondrites because they are dominantly composed (although some exceptions exist) of small spherules of once melted (igneous) rocks called chondrules.

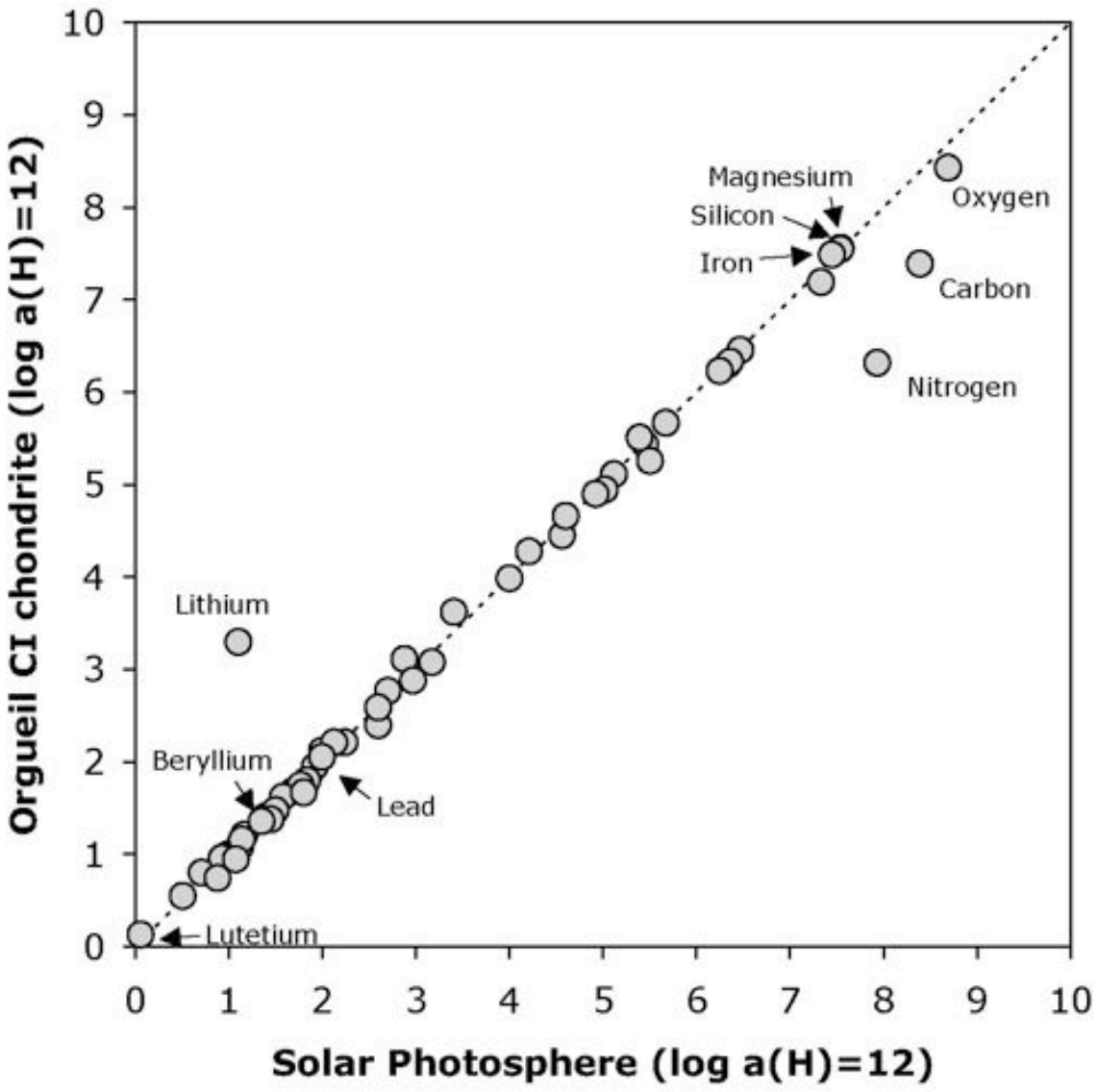

**Fig. 1**. Elemental abundances in CI meteorites and the solar photosphere. Data from Palme and Jones 2005.

A major property of chondrites is the strong similarity of their chemical composition with that of the solar photosphere (Fig. 1) for most elements except the most volatile ones (H, He, C, N and noble gases). This similarity indicates that chondrites were built from the initial solar system rocky material, without significant subsequent chemical fractionation. Small variations are observed in the proportions of the refractory vs volatile elements or of the siderophile (metal loving) vs lithophile (rock loving) elements, which are used to define families among chondrites (appendix 2). Chondrites have textures of aggregates of various nm- to mm-sized dust components and can be considered as cosmic sediments (Fig. 2). The major components of chondrites are (1) chondrules, which are ~mm-sized spherules of primordial solar system dust aggregates that were melted in the solar nebula and represent up to 80 vol% of chondrites. Other major components include (2) refractory inclusions, which formed at high temperature, are the oldest rocks of the solar system as indicated by isotopic ages of 4.567 Gyr (Amelin 2002) and range in size between a few μm and a few cm, (3) a fine-grained (nm- to μm-sized) unmelted matrix that cement all other components and (4) metal grains either associated with chondrules and refractory inclusions or isolated in the matrix.

Other samples useful to study the protosolar nebula include chondritic interplanetary dust particles produced during asteroid collisions or ejected from comets by the sublimation of their ices. Interplanetary dust particles are typically collected in the stratosphere and in polar regions, in which case they are commonly referred to as micrometeorites. Certified cometary materials are available owing to the recent Stardust mission, which brought back to Earth dust particles captured in the coma of the short-period comet 81P/Wild2. Finally, the Sun is a key object to study to understand the formation of the solar system, because it represents more than 99.8 % of the total mass of the solar system. Solar samples available for laboratory analyses are solar wind ions implanted in lunar soils from the Apollo and Luna missions or in the collectors of the Genesis mission, which collected solar wind samples during 2 years at the L1 Lagrange point between the Earth and the Sun.

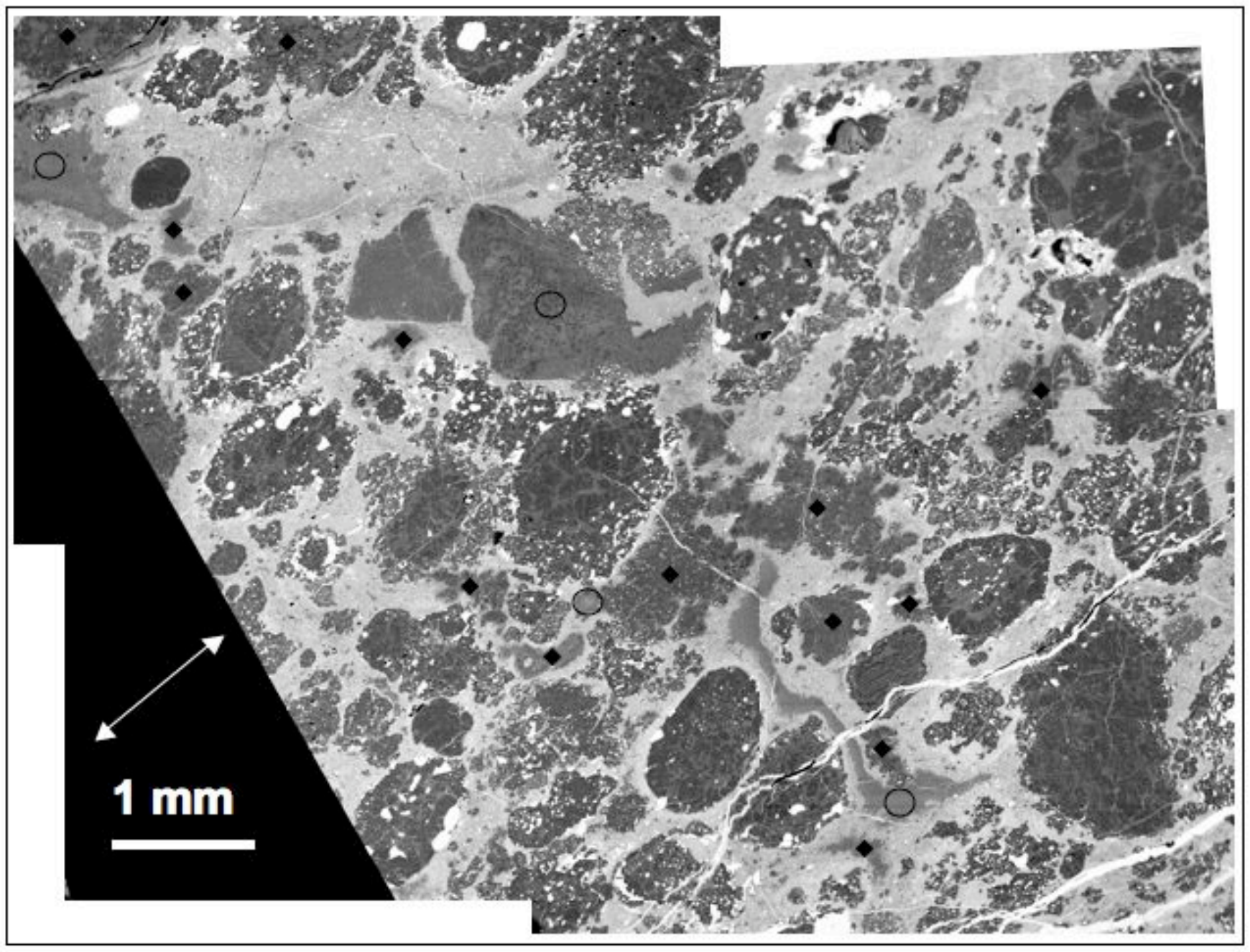

**Fig 2.** Typical texture of a carbonaceous chondrite : Efremovka. CAIs are located by empty circles, AOAs by full diamonds. Other dark grey objects are chondrules. The light grey interstitial material is the matrix. Note the preferential elongation direction (white arrow) due to a shock.

Here, several examples abundantly studied and still up to date are presented to show how the study of chondrites and their components can help to unravel the physics and the chemistry of the earliest stages of the solar nebula. These examples are (1) high temperature processes and the formation of chondrules and refractory inclusions, (2) the distribution of $^{16}O$ excesses and depletions among chondritic materials and their bearing on CO photochemistry and alternative models, (3) low temperature processes as revealed by organic chemistry and H isotopes, (4) evidence of irradiation by energetic particles from the young Sun as recorded by extinct radionuclides and extreme O isotope anomalies, and finally (5) homogeneities vs heterogeneties in the disk and the relationship between cometary and chondritic materials deduced from the study of interplanetary dust particles and cometary samples returned by the Stardust mission. Recent developments in solar system chronology are finally given in a short section. For each of these topics, a disk cartoon is given in appendix 1, which shows schematically the implications of meteorite studies for the physics and chemistry of the early solar system.

## 2. High temperature processes in the inner nebula

### *2.1. Chondrules*

#### *2.1.1. General petrographic description of chondrules*

Chondrules are 100 µm to 10 mm spherules of once melted silicates. They are dominated by ferromagnesian olivine and pyroxene minerals embedded in a Ca-Al-rich glassy mesostasis of close-to-feldspathic composition. Different groups of chondrules are recognized based on their textural and chemical properties (Fig. 3.). They can have porphyritic textures with coarse-grained olivine and pyroxene crystals, barred texture with skeletal crystals, radial textures with skeletal crystals radiating from the surface towards the interior, cryptocrystalline textures with micrometer sized crystals or they can be dominantly glassy with minor amounts of crystals. Based on their chemistry, two major groups are recognized : type I chondrules with reduced iron present as metal droplets or sulfides, together with magnesian olivine and pyroxene and type II chondrules with oxidized Fe entering the structure of the silicates.

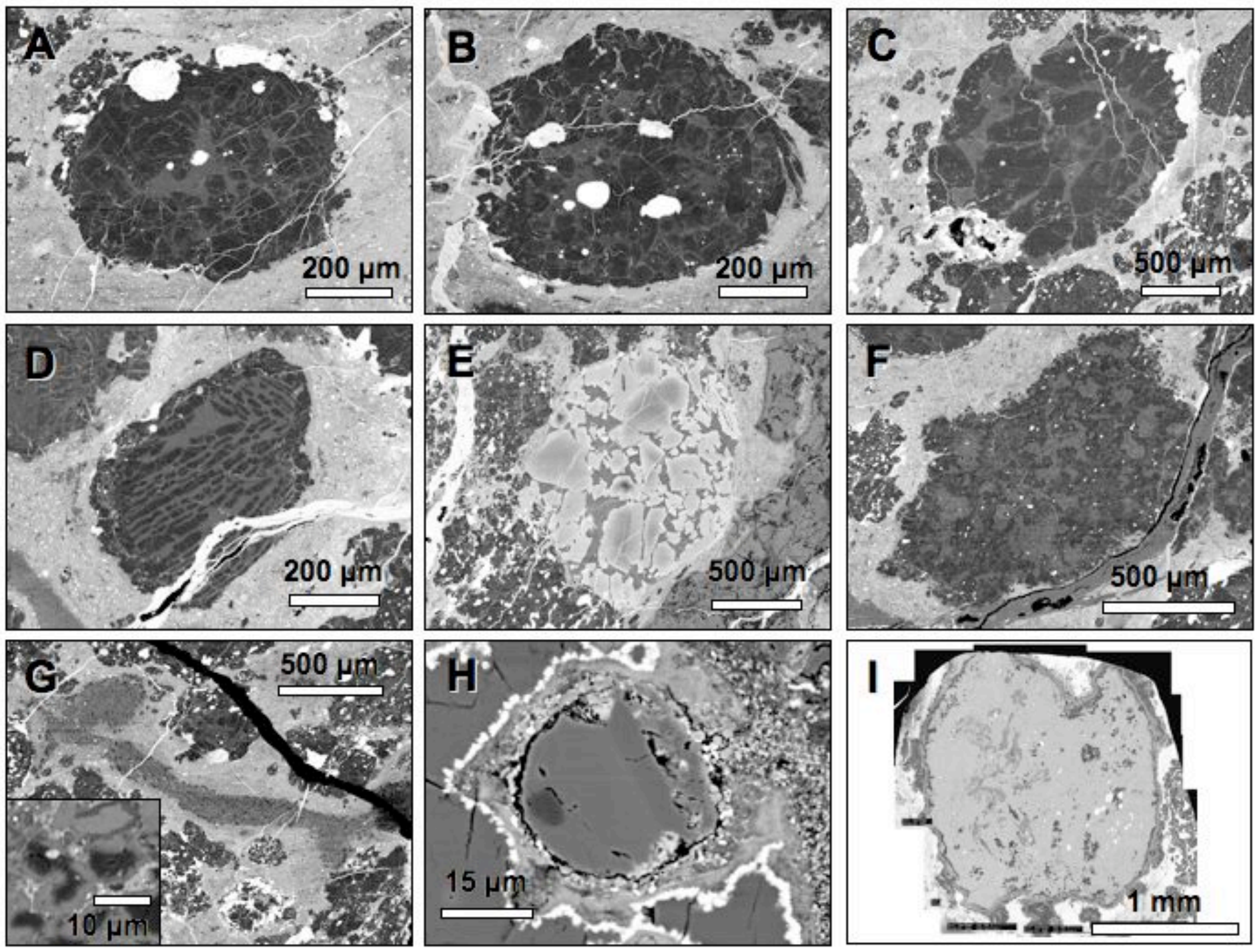

**Fig. 3**. Scanning electron micrographs of chondrules and CAIs. A. Porphyritic Olivine type I chondrule. B. Porphyritic Olivine Pyroxene type I chondrule. C. Porphyritic Pyroxene type I chondrule. D. Barred Olivine type I chondrule. E. Porphyritic Olivine type II chondrule. F. Amoeboid Olivine Aggregate. G. Fine-grained CAI with nodular texture shown in the insert. H. Hibonite-glass CAI spherule. I. Coarse-grained ultra-refractory igneous CAI E101,1 (ElGoresy et al. 2002). All objects are from the Efremovka carbonaceous chondrite, except H from the Murchison carbonaceous chondrite.

The most abundant chondrules are the type I porphyritic chondrules. Barred and radial chondrules are minor and glassy or cryptocrystalline chondrules are rare. A minor type of chondrules with more refractory chemistry intermediate between that of ferromagnesian chondrules and that of refractory inclusions, the Al-rich or anorthite-rich chondrules (ARC) are also present in chondrites but will not be described here. Porphyritic type I chondrules vary from olivine-rich (PO) to pyroxene-rich (PP) with intermediate members rich in olivine and pyroxene (POP). POP chondrules tend to have an olivine-rich core and a pyroxene-rich mantle. In POP and PP chondrules, olivine relicts are commonly found included in pyroxene.

### *2.1.2. The canonical model of chondrule formation*

Because chondrules are the major components of chondrites from which terrestrial planets formed, they have been abundantly studied (see reviews by Hewins et al. 1996, Zanda 2004, Sears 2004 and references therein) with a special emphasis on the porphyritic type I chondrules. Numerous experimental studies aiming at reproducing the observed textures in the laboratory have been perfomed in the past (e.g. crystallization, evaporation, oxydo-reduction) and chemical analyses of chondrules include most available elements and isotopes (major, minor and trace element abundances including alkalis and rare earth elements (REEs) and H, O, Mg, Si, S, K, Fe, Ni and noble gases isotopes for instance).

Among experimental studies, the crystallization of silicate melts of bulk chondrule chemical composition indicates that the various textures can be reproduced using variable maximum heating temperature and variable cooling rates (e.g. Radomsky and Hewins 1990). The presence of relict olivine suggests that the melting temperature was just below the liquidus of olivine. At the highest peak temperature (~1875 K), glassy chondrules are formed at high cooling rates (~1000 K/hour) and radial textures are obtained for lower cooling rates (down to 10 K/hour, e.g. Radomsky and Hewins 1990). For lower peak temperature (~1825 K), barred textures are obtained for high cooling rates and porphyritic textures are obtained at low cooling rates (e.g. Radomsky and Hewins 1990). Seeding experiments indicate that nucleation of crystals for a chondrule melt is eased with the presence of seed nuclei (Connolly and Hewins 1995). To achieve realistic porphyritic textures, coarse-grained seeds are however required, which resemble pre-existing porphyritic chondrules (Hewins and Fox 2004). The reproduction of porphyritic textures in the laboratory is thus not easy, except if the starting material has a porphyritic texture, which suggests an important recycling of chondrules.

Another important information is obtained from chemical and isotopic analysis of moderately volatile elements: alkalis (Na, K) for instance are present in chondrules in minor but significant abundance despite high temperature melting experiments indicate Na volatilization from silicate melts of orders of magnitudes in minutes (e.g. Tsuchiyama et al. 1981). In addition, Fe, Mg and especially K isotopes are not substantially fractionated by evaporative loss (e.g. Alexander et al. 2000), contrary to the expectations from high temperature melting in vacuum. The usual explanation for these contradictions is that chondrules were melted for a very short amount of time (flash melting in seconds to minutes) in a medium with a relatively high ambient pressure. High precision Mg isotope analysis indicates that the evaporation was reduced due to high ambient pressure ($P > 10^{-3}$ bars, Galy et al. 2000). The physical conditions advocated in the standard model of chondrule formation are summarized in Table 1.

**Table 1.** Physical conditions for standard chondrule formation

| | |
|---|---|
| Peak heating temperature | 1500-2000 K |
| Heating time | min - hours |
| Cooling rates | 10-100 K/ hours (porphyritic chondrules)<br>> 1000 K/hours (barred olivine chondrules) |
| Seed nuclei | Partial melting + repetitive events |

Many astrophysical environments and processes were investigated to explain the physical conditions required for chondrule formation: impacts, asteroidal volcanism, melting by lightning discharges, melting in the vicinity of planetesimals, melting by flares from the protosun or in jets and melting of pre-existing dust aggregates by shock waves in the disk (Hewins et al. 1996, Zanda 2004, Sears 2004 and references therein). Many of these processes are currently not favored. Detailed modelling of shock waves in the disk produces the best fit for the T and P profiles experienced by chondrules (e.g. Desch and Connolly 2002) so that the shock wave model is currently the favored hypothesis. Such shock waves are possibly driven by gravitational instabilities in the disk.

### *2.1.3. Chondrules as open-systems: a new vision*

Elemental mapping of chondrules by various techniques reveals that alkalis are enriched at the edge of the chondrules rather than in the core, which is the opposite to the pattern expected by partial evaporative loss during flash melting (e.g. Matsunami et al. 1993). Furthermore, condensation studies indicate that pyroxene can form from olivine by condensation of SiO in the vapour phase (Imae et al. 1993, Tissandier et al. 2002) following the reaction :

(1). $Mg_2SiO_4 + SiO + ½\ O_2 \rightarrow Mg_2Si_2O_6$

These observations suggest that moderately volatile elements such as alkalis and Si may have condensed from the gas into the chondrule melt. The additional observation that olivine in chondrules is not at equilibrium with the melt, indicates that olivine could not have formed from the melt but rather was a precursor mineral that reacted with $SiO_2$ to form pyroxene (Libourel et al. 2006). This hypothesis explains the mineralogical zoning of POP chondrules, the abundance of alkalis near the edge of chondrules and the presence of digested relict olivine within pyroxene. The O isotopic composition of chondrules is supportive of this model with an oxygen isotope evolution from olivine to pyroxene to mesostasis or silica in silica-bearing chondrules, towards the isotopic composition of the gas. Additionally, new observations of a compound CAI-chondrule object (Aléon and Bourot-Denise 2008) and of glassy aluminium-rich chondrules show spectacular chemical zoning supportive of this hypothesis (Nagahara et al. 2008).

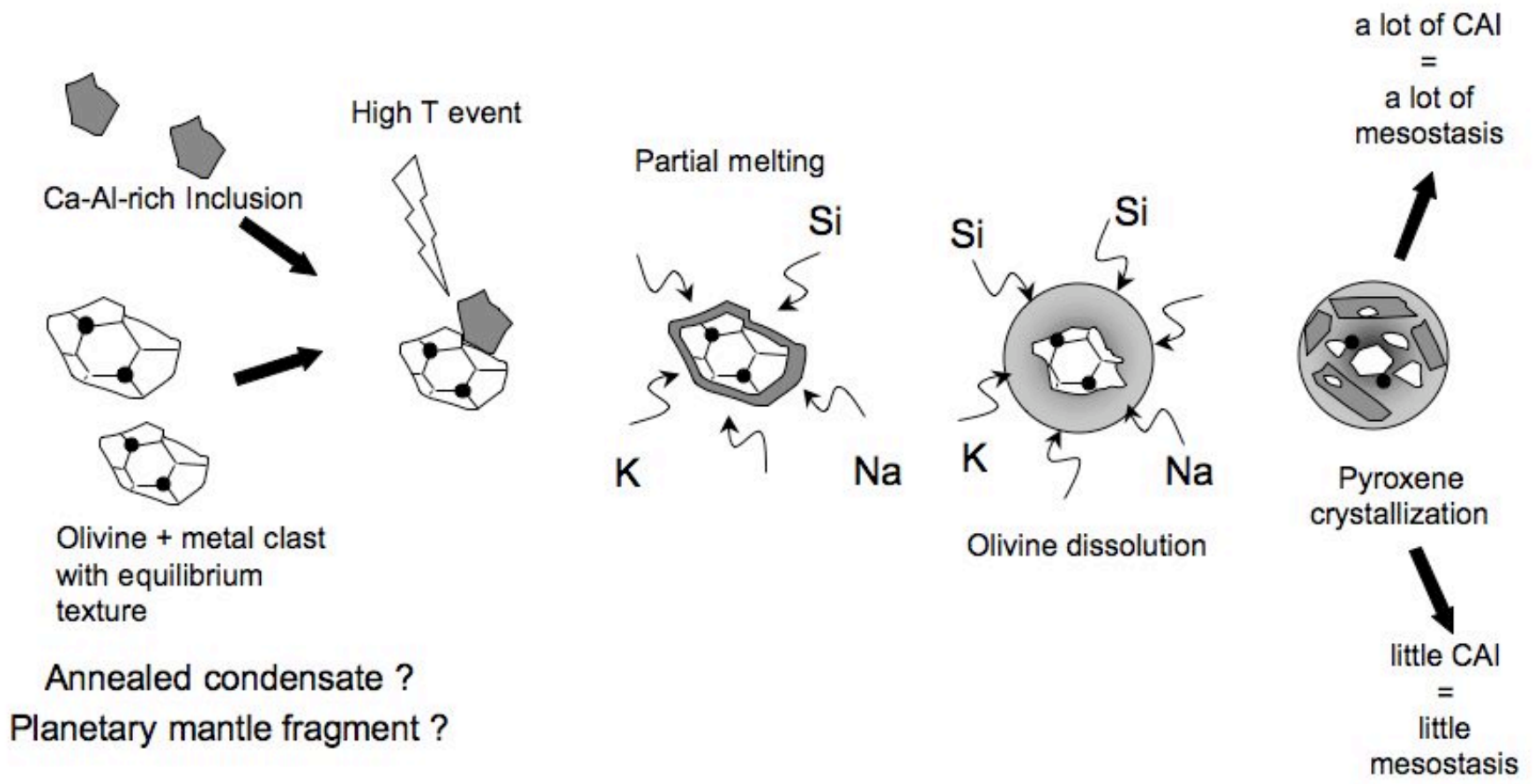

**Fig. 4.** Formation of chondrules as viewed from recent observations suggesting they behave as open systems with respect to the nebular gas (see text).

A major issue with this model is the origin of precursor olivine. Two models have currently been proposed. Based on the observation of equilibrium-growth textures with triple junctions (granoblastic textures) between olivine crystals in some chondrules, Libourel and Krot (2007) proposed that chondrules formed from olivine + metal aggregates, which crystallized slowly in a differentiated parent-body that was subsequently disrupted by a giant impact and later underwent melting in the nebula. By contrast, annealing experiments of olivine aggregates indicate that loose condensate aggregates could have developed triple junctions upon annealing in the nebula (Whattam et al. 2008). Typical porphyritic textures are produced by a subsequent melting episode in the presence of Ca-rich pyroxene and anorthite (Whattam and Hewins 2008). Regardless of the origin of the granoblastic olivines, the possibility that chondrules were an open-system with relict crystals reacting with a melt of changing composition because moderately volatile elements condensed from the gas into the melt (Fig. 4.), implies that peak heating temperatures, cooling rates and densities in the gas could be different from those expected by the shock wave model.

## *2.2. Refractory inclusions*

### *2.2.1. General petrographic properties of refractory inclusions*

Refractory inclusions are the oldest objects of the solar system with a Pb isotope age of 4.567 Gyr (Amelin et al. 2002, Connelly et al. 2008). They include very different objects made of the same high temperature minerals: Ca-Al-Ti-rich oxides, Ca-Al-Ti-rich silicates and Mg-rich olivine (forsterite) (appendix 3). There is a wide variety of refractory inclusions, which can be grouped in roughly three families (Fig. 3): (1) coarse-grained once molten Ca-Al-rich inclusions (CAIs) with magmatic textures and crystal size ranging from ~10 µm to ~1 mm, (2) fine-grained CAIs consisting of aggregates of 20-50 µm nodules with concentric mineral zoning and µm to sub-µm crystal size, and (3) amoeboid olivine aggregates (AOAs), which are amoeboid shaped inclusions consisting of fine-grained CAI nodules and veins in a matrix of forsteritic olivine ± FeNi metal.

### *2.2.2. Refractory inclusions formed by condensation*

Calculated mineral equilibrium stability fields indicate that CAI minerals formed by condensation of a gas of solar composition at high temperature > 1200 K (e.g. Grossman 1972, Yoneda and Grossma 1995, Petaev and Wood 1998, Ebel and Grossman 2000). Results of equilibrium condensation calculations indicate a satisfactory despite variable fit with CAI mineralogy for pressures in the $10^{-6} - 10^{-1}$ bars range (e.g. Ebel and Grossman 2000). A recent study on 17 igneous CAIs (Grossman et al. 2008) confirms that the precursors of these CAIs could have formed by condensation at different pressures in this range, from either a gas of solar composition or from a gas enriched by a factor 10 in rock-forming elements from pre-evaporated dust. A theoretical condensation sequence has been established based on condensation temperatures of the various minerals at thermodynamical equilibrium (Grossman 1972). In the theoretical condensation sequence only corundum, perovskite, olivine and FeNi metal form by direct condensation from the gas, all other phases are formed by subsequent reactions of these starting minerals with the gas. Until recently, only theoretical calculations were available to establish the condensation origin of CAI and AOA minerals. These minerals have been successfully produced by laboratory condensation of a gas of solar interelemental ratios Si/Mg/Ca/Al at $10^{-3}$ bars in 4 min to 1 hour (Toppani et al. 2006a). In this experiment, crystalline silicates and oxides were produced both by homogeneous nucleation and aggregation in the gas and by heterogeneous nucleation on the collection plate. An experimental condensation sequence has been defined that is remarkably close to the theoretical sequence, given the experimental limitations (Toppani et al. 2006a). This experimental condensation sequence suggests that minor discrepancies between the theoretical sequence and CAI observations may be due to kinetic effects. Grossman et al. (2008) indicate that the range of pressures estimated for condensation of CG-CAI precursors broadly agrees with dynamic modelling of disk midplane regions having a temperature in the 1000 K - 2000 K range.

Fine-grained CAIs and AOAs are usually thought to be aggregates of direct gas-solid condensates. Nodules in unaltered fine-grained CAIs have mineralogical zoning that follow the condensation sequence, with the most refractory minerals in the core (e.g. Krot et al. 2004). They have REE abundances that are volatility controlled and indicate condensation from a reservoir already depleted in the most refractory REEs due to a first episode of condensation in the nebula (Boynton 1975, Grossman and Ganapathy 1976). By contrast, coarse-grained CAIs probably formed from condensates that subsequently underwent important partial melting. They are commonly surrounded by a rim of fine-grained minerals analogous to those of fine-grained CAIs, with the most refractory minerals towards the interior of the CAI, the so-called Wark-Lovering rims (Wark and Lovering, 1977). These rims are commonly attributed to a late episode of condensation in the nebula (Wark and Lovering, 1977). However, a recent nanometre scale mineralogical study suggests that they may have formed from a highly refractory melt, which reacted with the gas (Toppani et al. 2006b).

### *2.2.3. Coarse-grained CAIs as probes of nebular conditions*

Coarse-grained CAIs (CG-CAIs) have magmatic textures that indicate crystallization from a melt. This has been clearly demonstrated by crystallization experiments and theoretical calculations for the most commonly studied CG-CAIs, the type B CAIs from Allende and other CV3 chondrites. The experimental crystallization of silicate melts of type B CAI composition reveal a sequence of crystallization similar to that observed in natural CAIs : spinel crystallized at ~1800 K followed by melilite, followed by clinopyroxene, followed by anorthite with decreasing temperature (Stopler 1982). In addition, petrologic evidence in natural CAIs suggest peak heating temperature close to 1700-1800 K. However, the chemical composition of bulk CG-CAIs does not match exactly that predicted by the melting and crystallization of condensates but is better explained by condensation plus an additional loss of Si and Mg by evaporation (Grossman et al. 2002, 2008). Similarly, Si and Mg isotopes are commonly enriched in the heavy isotopes, indicating a loss of the lighter isotopes by distillation upon evaporation (Richter et al. 2002, Grossman et al. 2008). Modelling of the isotopic fractionation as a function of pressure of $H_2$ indicates that type B CAI melts underwent evaporation at pressures around $10^{-6}$ bars (Richter et al. 2002).

Crystallization experiments at various cooling rates indicate that the textures and crystal sizes of natural CAIs are best matched by slow cooling rates (0.1 to 25 K/hour, Stolper and Paque, 1986). The isotopic fractionation of Mg measured in CAIs furthermore restricts this range to 1 to 10 K/hour (Richter et al. 2002). These cooling rates are slow enough to be controlled by the cooling of the external medium rather than by radiative cooling in a cold environment. They are too rapid for cooling in a dense region at the disk mid-plane where cooling rates << 1 K/hour are expected and too slow for cooling in an optically thin region where cooling is expected to proceed faster than 100 K/hour (Stolper and Paque 1986, Richter et al. 2002). It has thus been proposed that CG-CAIs underwent high temperature melting in a wind close to the protosun, where they were evaporated at low ambient pressures and underwent slow cooling when transported to the accretion regions of chondrites and terrestrial planets at a few AU (Richter et al. 2002, Shu et al. 1996).

## 3. Oxygen isotopes and the origin of the planet building blocks.

### *3.1. General properties of oxygen*

Oxygen is the third most abundant element in the galaxy after H and He. The recently revised abundances of O and C in the Sun are 8.66 dex (Asplund et al. 2004) and 8.39 dex (Asplund et al. 2005), respectively (given as log aO and log aC with log aH = 12), resulting in a C/O ratio of the Sun (i.e. of the nebula) of 0.54. O is thus a major element in the gas phase, present dominantly as CO, the most abundant gas-phase molecule after $H_2$, and $H_2O$ and to a lesser extent as SiO. It is also the major element in the dust, being the major constituent of silicates and oxides, the most abundant rock-forming minerals.

Oxygen has three stable isotopes synthesized in stars. $^{16}O$ is a primary isotope formed by the fusion of four He nuclei during He burning in supernovae, whereas $^{17}O$ and $^{18}O$ are secondary isotopes formed from pre-existing $^{16}O$ and $^{14}N$, respectively, during H burning in the CNO cycle of low to intermediate mass stars and under a limited range of temperatures in the He burning shell of massive stars, respectively. This difference in nucleosynthesis results in drastically different abundances in the solar system, with $^{16}O$ being approximately 500 times more abundant than $^{18}O$ and 2500 times more abundant than $^{17}O$.

With the exception of a small amount of exotic grains present in the matrices of chondrites, which are believed to have condensed in the atmospheres of evolved stars before the birth of the solar system (and are thus referred to as presolar interstellar grains) and which preserved large isotopic variations directly attributable to nucleosynthetic processes, oxygen isotopic variations in the solar system are limited to a few %. Most O isotopic ratios are thus reported in delta notation following the convention from terrestrial geochemistry. They are given as relative deviations from a well established reference, in parts par mil (e.g. eqn. 2. for the $^{18}O/^{16}O$ ratio). The reference is the Standard Mean Ocean Water (SMOW) and has $^{18}O/^{16}O$ ratios and $^{17}O/^{16}O$ ratios of $2.0052 \times 10^{-3}$ and $3.8288 \times 10^{-4}$, respectively.

(2) $\delta^{18}O = [(^{18}O/^{16}O_{sple} - {}^{18}O/^{16}O_{ref}) / {}^{18}O/^{16}O_{ref}] \times 1000$

Most chemical and physical processes are characterized by an isotopic fractionation between the products and the reactants, which depends on the relative masses of the isotopes. A fractionation factor $\alpha$ is defined as the ratio of the final isotopic ratio to the initial isotopic ratio (e.g. for a mineral crystallized from a magma $\alpha_{mineral-magma} = {}^{18}O/^{16}O_{mineral} / {}^{18}O/^{16}O_{magma}$). The dependence of $\alpha$ to the masses is best given by an exponential law, which can be approximated by a linear law (eqn. 3) of sufficient precision for most studies of meteorites.

(3) $\alpha = 1 + (M2-M1) \times \delta$ with $\delta \ll 1$

where M2 and M1 are the exact masses of the isotopes [e.g. masses of $^{18}O$ and $^{16}O$ in the case of the $^{18}O/^{16}O$ ratio] and $\delta$ is the fractionation intrinsic to the process.

O isotope variations in meteorites are commonly reported in a graph where $\delta^{17}O$ is shown as a function of $\delta^{18}O$ (Fig. 5). In this graph, mass dependant fractionations plot on a line of slope 0.52, which depends directly on the relative masses of $^{16}O$, $^{17}O$ and $^{18}O$. The Terrestrial Mass Fractionation (TMF) line is by definition the mass fractionation line that goes through 0. Mass balance calculations show that any mixture of two components plot along a straight line of any slope called a mixing line. An additional parameter is introduced, $\Delta^{17}O$, which defines the $\delta^{17}O$ deviation between a given composition and the TMF line (Eqn 4).

(4) $\Delta^{17}O = \delta^{17}O - 0.52 \times \delta^{18}O$

Note that isotopic fractionations depend on temperature. For instance, the isotopic fractionation between two minerals crystallized from the same magma can be used to determine the temperature of crystallization of this magma.

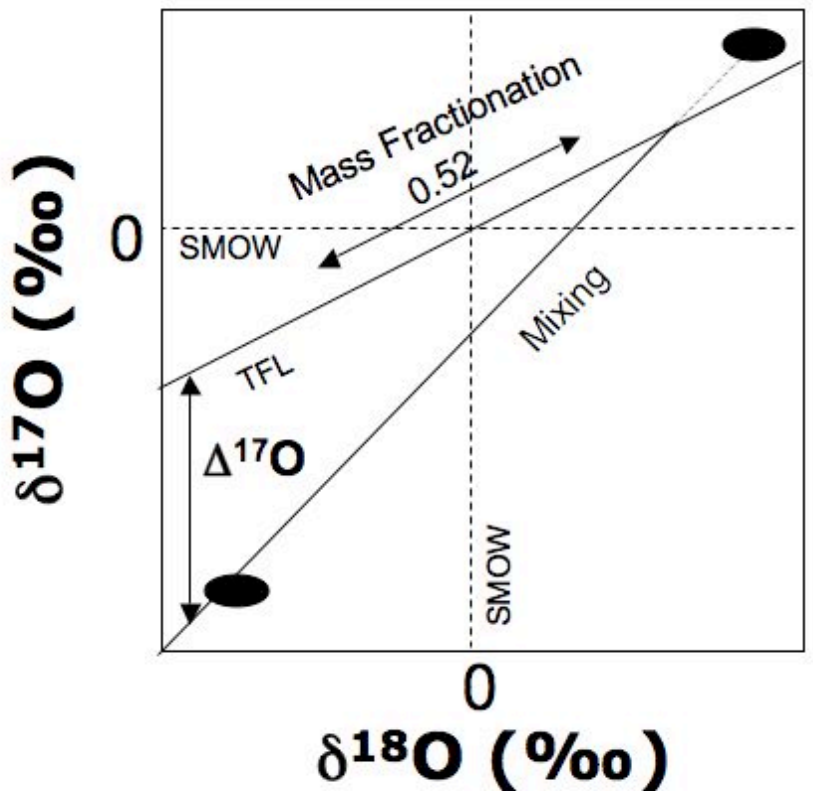

**Fig. 5.** Principles of the three oxygen isotope plot (see text). Dashed lines indicate the terrestrial composition. TFL stands for Terrestrial mass Fractionation Line. SMOW stands for Standard Mean Ocean Water, the terrestrial oxygen isotope reference.

## *3.2. The oxygen isotope anomaly in meteorites*

### *3.2.1. Historical perspectives*

1969 is a key year for the study of early solar system and planet formation : Man walked on the Moon on july 16th and starting from the Apollo XI mission, six missions brought back to Earth about 382 kg of lunar rocks. However, it is much less known that earlier that year, on february 8th, a meteorite shower of more than a 1000 stones fell in the village of Pueblo de Allende in Mexico, accounting for more than 2 tons of the largest carbonaceous chondrite ever recovered. During the following years, Robert Clayton and his team at the university of Chicago, measured O isotopes in lunar rocks and magmatic inclusions (i.e. CAIs and chondrules) from Allende to determine their temperature of crystallization. Their work yielded a very important discovery: when the three oxygen isotopes are measured, the Allende samples spread along a slope ~1 (exact slope is 0.94) in the three isotope diagram, which can be accounted by the admixture of pure $^{16}O$ (Clayton et al. 1973) and was later referred to as the Carbonaceous Chondrite Anhydrous Minerals mixing line (Clayton and Mayeda 1984). Almost 40 years later, numerous analyses have shown that all planetary materials display this variation in $^{16}O$ content at various levels (Fig 6) but the origin of this isotopic anomaly is still unclear. Understanding the origin of oxygen isotope variations and distributions in meteorites is thus a fundamental issue to understand the origin of planetary materials and their genetic relationships.

Bulk analyses have shown that the oxygen isotopic composition of meteorites can be roughly explained by a limited number of processes : (1) mixture in various amounts of components with different $^{16}O$ content along the slope 1 line, (2) mixture with a mass fractionated component formed by aqueous alteration at low temperature on a parent-body (e.g. hydrated carbonaceous chondrites) and (3) mass fractionation during metamorphism (e.g. ordinary chondrites) or geologic processes (planetary differentiates). Therefore, the main issue to unravel the origin of this $^{16}O$ variation is to characterize primordial nebular components that escaped parent-body processing and understand their formation.

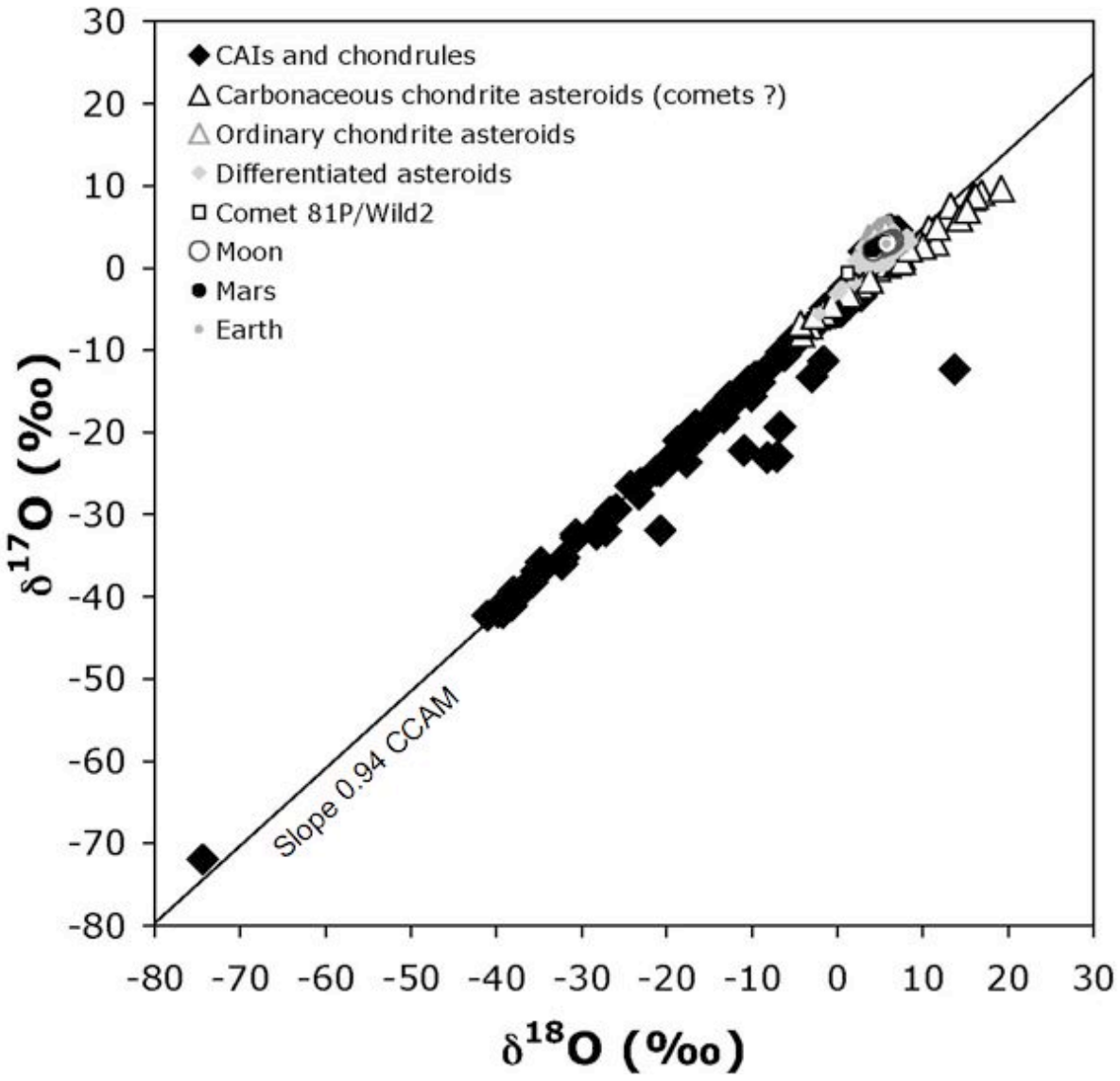

**Fig. 6**. Oxygen isotopic composition of planetary materials in the solar system. All samples plot along a ~slope 1 line. Minor departures from this line are due to mass fractionation upon evaporation (Fractionated CAIs) and hydration (hydrated carbonaceous chondrites). The Earth (average mantle), Mars and the Moon plot almost on top of each other at this scale. With exception of the lowermost chondrule and of comet Wild2, all data are from a compilation of oxygen isotope data from the Chicago group (compilation G.J. MacPherson and A.M. Davis, data kindly provided by R.N. Clayton and A.M. Davis). The lowermost chondrule datapoint is an average of analyses by Kobayashi et al. (2003) and the datapoint for comet Wild2 is an average of olivine and pyroxene analyzed by McKeegan et al. (2006).

### *3.2.2. Oxygen isotopes in CAIs and chondrules*

The largest $^{16}O$ excesses are commonly found in the least melted CAIs. The oxygen isotopic composition of CAIs (Fig. 7a) indicate mixture between (i) an endmember having a 5% excess $^{16}O$ relative to the Earth ($\delta^{18}O \approx -50‰$; $\delta^{17}O \approx -50‰$; $\Delta^{17}O \approx -25‰$), attributed to the isotopic composition of the CAI precursors, that is the composition of the gas from which they condensed (Krot et al. 2002) and (ii) an endmember of close to terrestrial composition resulting from a late isotopic exchange during melting with a gaseous reservoir of close-to-terrestrial isotopic composition (e.g. Yurimoto et al. 1998), hereafter referred to as the planetary gas. The detailed mechanism of isotopic exchange between CAIs and the planetary gas are still not fully understood. Indeed, the degrees of $^{16}O$ enhancements in minerals from igneous CAIs agree better with the self-diffusivities of O at the solid state in these minerals than with their crystallization sequence (Ryerson and McKeegan 1994). However, the crystal sizes are so that self-diffusion timescales do not support a solid state isotopic exchange after CAI final solidification (Ryerson and McKeegan 1994).

In chondrules, (Fig. 7b) oxygen isotope variations are mostly restricted to about 1-2% close to the terrestrial composition. Bulk chondrule analyses define compositional trends deviating from the slope 0.94 line. Chondrules from carbonaceous chondrites have variable $^{16}O$ excesses whereas chondrules from ordinary chondrites have small $^{16}O$ depletions (e.g. Gooding et al. 1983, Clayton et al. 1983, 1991). Chondrules from enstatite chondrites are mass fractionated along the TMF line (Clayton et al. 1984). These compositional trends meet at the intercept with the terrestrial composition. Inter-mineral oxygen isotopic variations in porphyritic chondrules from carbonaceous chondrites have recently been interpreted as resulting from isotopic exchange between an olivine-rich precursor having variable $^{16}O$ excesses and the planetary gas during melting (Chaussidon et al. 2008). Minerals crystallized from the melt (pyroxene) have intermediate isotopic composition, while the residual mesostasis is closest to the gas in isotopic composition. These observations support formation of chondrules by open-system melting with condensation of moderately volatile elements from the gas (Libourel et al. 2006). From mass balance calculations, the isotopic composition of the planetary gas has been found to be $\delta^{18}O$= 3.6 ± 1‰, $\delta^{17}O$=1.8 ± 1‰, at the intercept between the trends of carbonaceous chondrite chondrules and ordinary chondrite chondrules and barely mass fractionated relative to the enstatite chondrite chondrules and the mantle of the Earth and Mars (Chaussidon et al. 2008). Rare unusual chondrules have been found to have large isotopic heterogeneities with $^{16}O$ excesses up to 8% (Jones et al. 2004, Kobayashi et al. 2003).

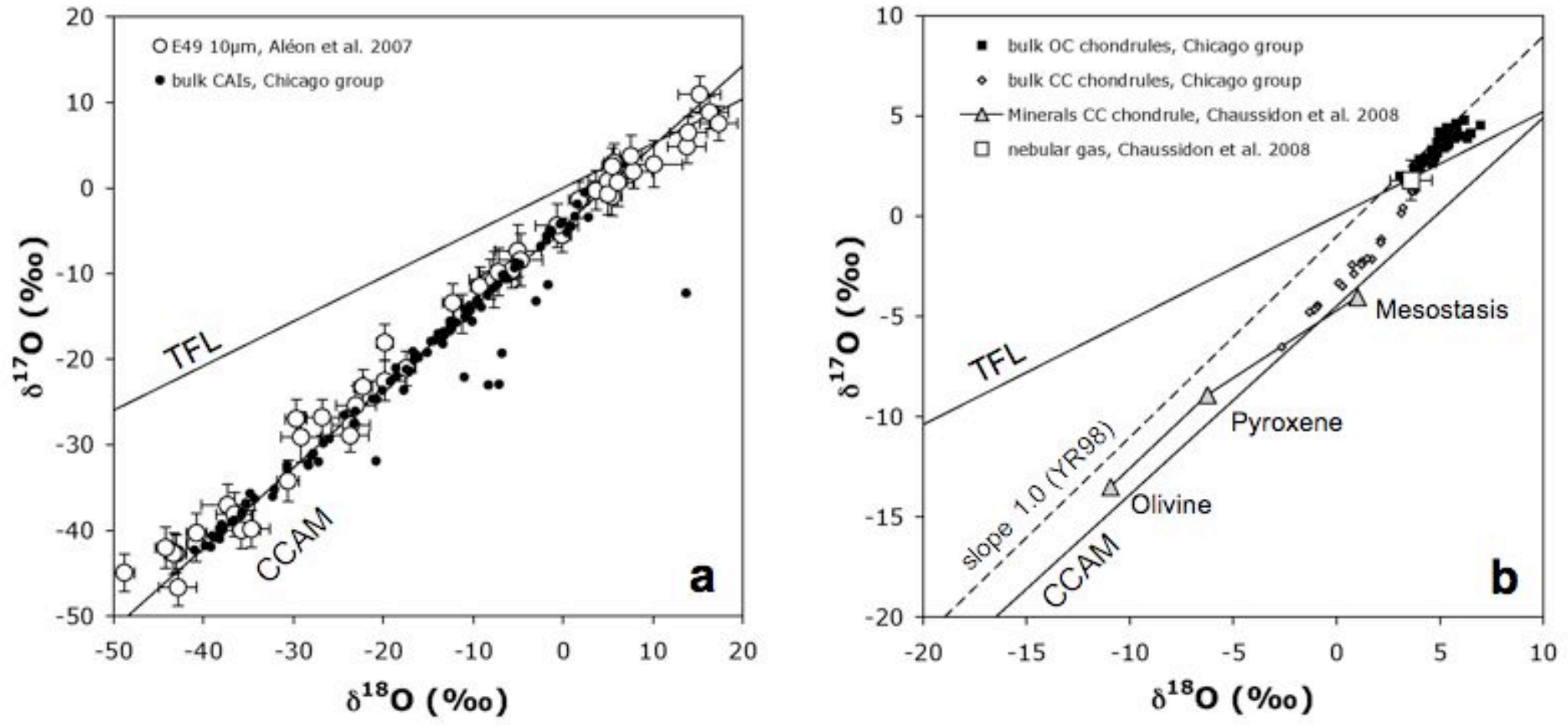

**Fig. 7**. Oxygen isotope distribution in CAIs (a) and chondrules (b). Bulk analyses are from the compilation of the data from the Chicago group (see fig. 6). Other datapoints are individual minerals from a single CAI (a) that cover the whole range of oxygen isotopic composition in CAIs (Aléon et al. 2007) and averages of individual mineral analyses in a single chondrule (b) that depicts the oxygen isotope evolution from olivine towards the nebular gas during chondrule crystallization (Chaussidon et al. 2008). TFL is the Terrestrial mass Fractionation Line, CCAM is the slope 0.94 line observed in Carbonaceous Chondrites Anhydrous Minerals and YR98 is the slope 1.0 line predicted to represent the primordial trend before CAI/chondrule late processing (Young and Russell 1998).

These observations suggest that CAI and chondrule precursors formed in a gaseous reservoir enriched in $^{16}O$ and were later remelted in a gas of isotopic composition similar to that of terrestrial planets, which implies a large scale change of O isotopic composition in the nebular gas during planet formation. Whether this change of isotopic composition is spatial or temporal strongly depends on the models advocated to explain the origin of the $^{16}O$ anomaly. This anomaly appears to be preserved in planet-sized objects, although at a much smaller scale, making it a powerful tracer of planet formation. For instance, Mars appears to be depleted in $^{16}O$ by 0.321 ± 0.014‰ relative to the Earth (Franchi et al. 1999), while differentiated meteorites expected to come from the asteroid 4-Vesta are enriched in $^{16}O$ by 0.239 ± 0.007‰ (Greenwood et al. 2005; Fig 8.).

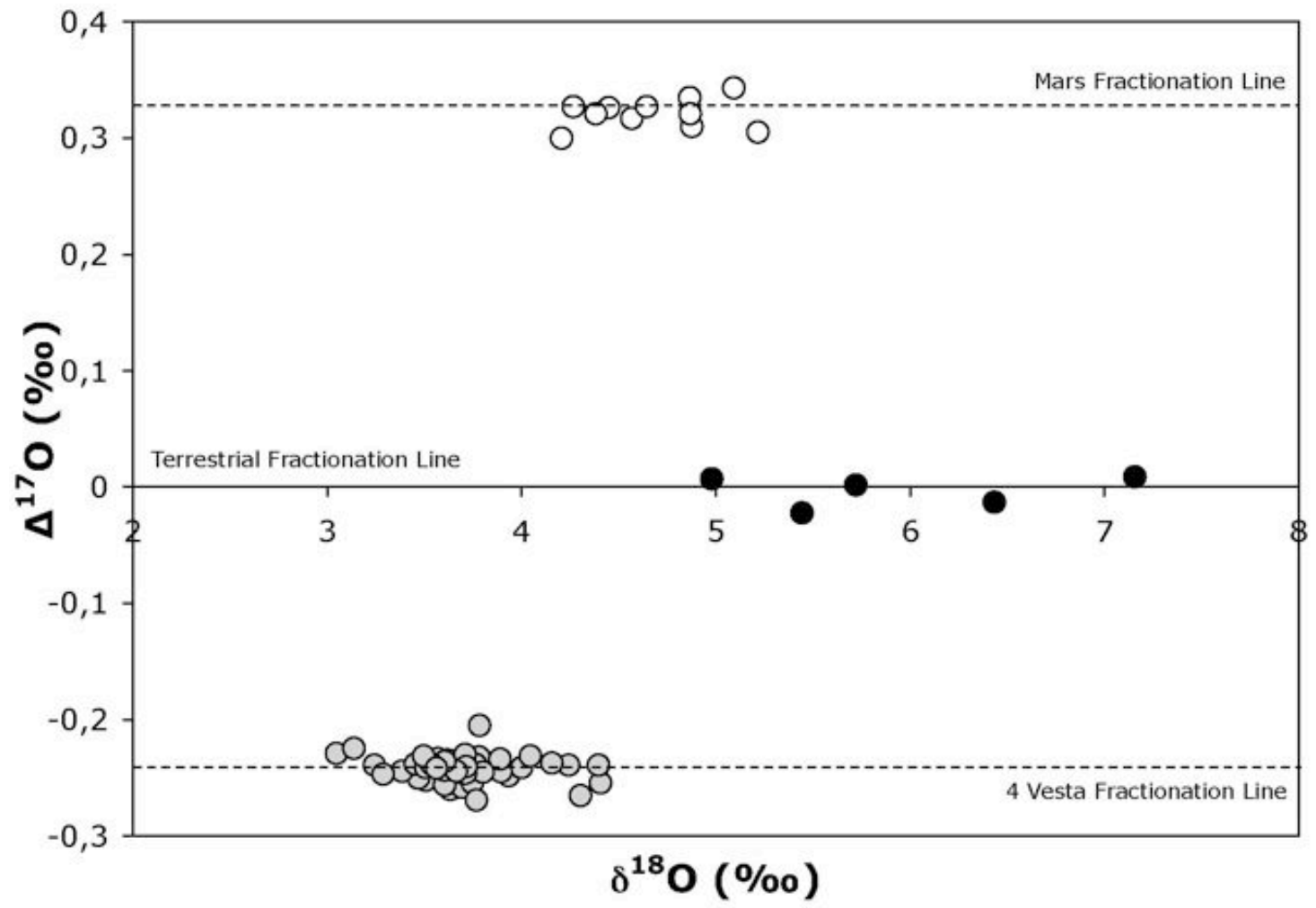

**Fig. 8**. Oxygen isotopic composition of planet sized objects. The Mars composition is from high precision analyses of martian meteorites (Franchi et al. 1999) and the 4Vesta composition is from high precision analyses of Howardites, Eucrites and Diogenites (HED) meteorites by Greenwood et al. (2005).

### *3.3. Origin of the oxygen isotope anomaly*

Several models have been advocated to explain the origin of the oxygen isotopic anomaly in the early solar system. The initial proposition was injection of pure nucleosynthetic $^{16}O$ from a nearby supernova into the collapsing presolar molecular cloud (Clayton et al. 1973, 1977). However, this hypothesis is now discarded. Indeed, the $^{16}O$ excess is not correlated with excesses in He-, C- and O-burning isotopes of other major elements (e.g. $^{24}Mg$ or $^{28}Si$) typical of type II supernovae. In addition, presolar interstellar oxide and silicate grains were discovered in 1994 and 2003, respectively (Nittler et al. 1994, Messenger et al. 2003, Nguyen and Zinner 2004, Nagashima et al. 2004). Their extensive study (e.g. Nittler et al. 1997, Mostefaoui et al. 2004, Nguyen et al. 2007) has shown that most O-bearing interstellar dust grains injected in the early solar system are $^{17}O$-rich and $^{18}O$-poor indicating a source from RGB (red giant branch) and AGB (asymptotic giant branch) stars, in agreement with these stars being major producers of dust in the galaxy (e.g. Jones 1997).

Thiemens and Heidenreich (1983) reported laboratory experiments resulting in preferential $^{16}O$ fractionation due to isotope selective photodissociation of $O_2$ upon self-shielding of the $^{16}O^{16}O$ isotopomer. They proposed that this mechanism could have occurred in the early solar system, resulting in the isotopic anomaly observed in CAIs and chondrules. Despite the self-shielding of $O_2$ was later ruled out by Navon and Wasserburg (1985), this study opened the way to the now-accepted idea that the slope-1 line in meteorites is created by a major chemical process occurring in the solar nebula. This process is, however, still to be firmly identified and two categories of mechanisms have been proposed : non-mass dependant isotopic fractionation during chemical reactions or isotope selective photodissociation of CO upon self-shielding of $C^{16}O$.

### *3.3.1. Non mass dependant chemistry*

Non-mass dependant isotopic fractionations created during ozone chemistry in the stratosphere result in more than 10% excesses in $^{17}O$ and $^{18}O$ observed in the $O_3$ molecule (e.g. Mauersberger 1987, Schueler et al. 1990). The mechanism responsible for these fractionations has been studied in detail (e.g. Thiemens 1999, 2006) and two explanations have been proposed. The most accepted model is based on a quantum mechanical effect resulting in reaction rates differences between symmetric and non-symmetric isotopomers. This quantum effect has been investigated in detail for ozone chemistry and the model reproduces fairly well the observed relative abundances of the various ozone isotopomers (Hathorn and Marcus 1999). A surface chemistry version of this model has been developed to account for the preferential incorporation of $^{16}O$ in a growing CAI (Marcus 2004).

An alternative model has been proposed to explain observations in both ozone (Robert and Cami-Perret 2001) and meteorites (Robert 2004). This model is based on reaction rates differences between distinguishable isotopes and non-distinguishable isotopes during reactive collisions. Again, this model reproduces fairly well the observed isotopic abundances in stratospheric ozone (Robert and Cami-Perret 2001).

These two models have common implications and differences. Both require gas-phase chemistry during the condensation of solids and in both cases the chemical reaction(s) responsible for the isotopic anomaly in the conditions of the solar nebula is (are) still unknown. However, whereas the molecular symmetry model implies that only O and potentially S can show such non-mass dependant effects due to their peculiar bonding chemistry, the distinguishability model could affect many elements. In that respect, Robert (2004) shows that besides O isotope anomalies, many isotopic anomalies in meteorites previously attributed to stellar nucleosynthesis (Xe, Ti, Mo, Sm, Si, Ba, Nd) can be explained by the distinguishability between isotopes during reactive collisions.

### *3.3.2. Isotope selective photodissociation of CO : self-shielding*

CO, the major gas phase O-bearing molecule in the nebula, is photodissociated by UV light by a predissociation mechanism. Because $C^{16}O$ is much more abundant than the other isotopomers, when the optical thickness is large enough, photons in the 91-110 nm range with energies corresponding to its rovibrational lines are absorbed and $C^{16}O$ stops being photodissociated. By constrast, the photons with energies corresponding to the rovibrational lines of the other isotopomers are never totally absorbed due to the much lower abundances of the latter and $C^{17}O$ and $C^{18}O$ are still photodissociated. Due to this isotope selective photodissociation, the resulting gas becomes preferentially enriched in $C^{16}O$ relatively to the other isotopomers whereas atomic oxygen becomes enriched in $^{17}O$ and $^{18}O$. This mechanism has been investigated in details for interstellar clouds (e.g. Van Dishoeck and Black 1988, Warin et al. 1996), where it is known that self-shielding occurs and astronomical observations have revealed large isotopic shifts in CO due to self-shielding around protostars (e.g. Sheffer et al. 2002, Federman et al. 2003).

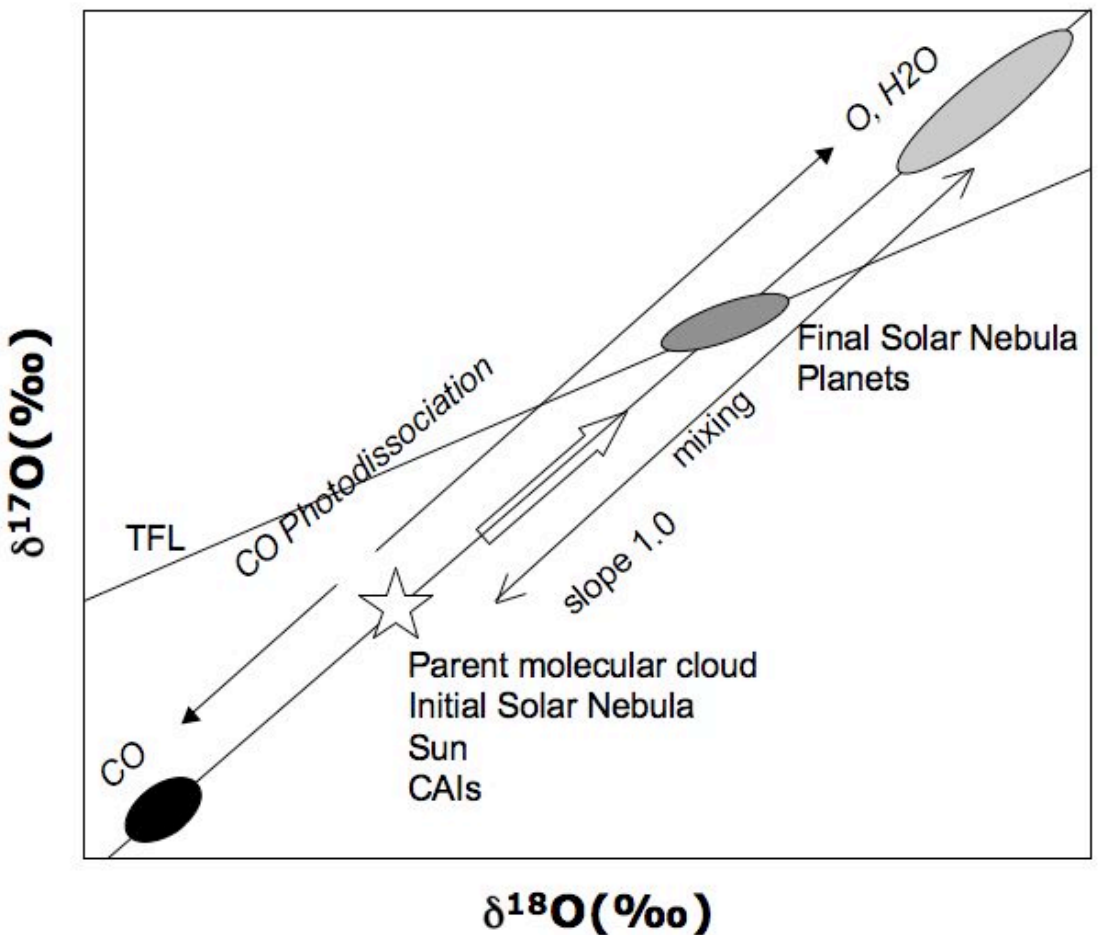

**Fig. 9**. Evolution of oxygen isotopes in the early solar nebula expected in the self-shielding models. The starting composition is indicated by a star. From this composition CO becomes enriched in $^{16}O$, while atomic O and water become enriched in $^{17}O$ and $^{18}O$ creating a slope 1 line. Rocky materials subsequently evolve from the starting composition towards the planetary composition by isotope exchange with water, that is along the slope 1 line.

CO self-shielding can explain the isotopic anomaly in meteorites if $^{17}O$ and $^{18}O$ produced by the selective photodissociation of the minor CO isotopomers can be efficiently transferred to $H_2O$ by reaction with hydrogen. An efficient transport of $H_2O$ inside the snow line would ensure evaporation and transfer of the anomaly to the gas which would become heavier. Isotopic exchange with silicates is much more efficient with $H_2O$ than with CO (Yu et al. 1995, Boesenberg et al. 2005). This would result in the preferential enrichment of the dust in heavy O isotopes. The self-shielding hypothesis thus imply that the starting solar system composition was $^{16}O$-rich similarly to CAIs and became progressively $^{16}O$-poor as accretion of $^{16}O$-poor ices proceeded (Fig. 9.). In that respect, CAIs could be considered "isotopically normal" while planets (and life) would be "isotopically anomalous" (Clayton 2002).

Three self-shielding models have been investigated to explain the O isotopic anomaly in meteorites, which differ on the location of the photodissociation. Clayton (2002) proposed that self-shielding occurred at the X-point, the inner edge of the accretion disk in the framework of the protoplanetary disk model developed by Shu et al. (1996). This model implies that all the matter from which the terrestrial planets are made was somehow recycled through this extremely localized region of the nebula. It remains to be physically and chemically investigated in detail. The second model proposed that self-shielding occurred in the parent molecular cloud of the Sun and that interstellar $^{16}O$-poor water ice was injected in the disk and transported inwards during accretion until it reached the snow line where water ice would be vaporized resulting in the $^{16}O$ depletion of the gas (Yurimoto and Kuramoto 2004, Lee et al. 2008). The third model proposed that self-shielding occurred at the surface of the outer disk (5 AU – 100 AU) upon irradiation by both interstellar UV photons and UV photons from the protosun (Lyons and Young 2005, Young 2007). $^{16}O$-poor water formed at the surface of the disk would freeze and settle onto the midplane before being transported into the inner regions of the disk. Timescales for isotopic change in the inner nebula deduced from chemical and transport models (Lyons and Young, 2005, Young 2007) are in the $10^5$-$10^6$ years range.

Recent support was brought to the self-shielding models, first by the discovery in the primitive carbonaceous chondrite Acfer 094 of large $^{16}O$-depletions (up to 20%) in an intergrowth of Fe-sulfides (pyrrhotite) and Fe-oxides (magnetite), nicknamed "cosmic symplectite" (COS), the formation of magnetite being attributed to oxidation of metal by $^{16}O$-poor water below 360 K (Sakamoto et al. 2007, Seto et al. 2008). The second support to the self-shielding model came from the determination of the O isotopic composition of the Sun by the Genesis mission (McKeegan et al. 2008). Two previous estimates of the Sun composition from solar wind implanted in lunar metal grains gave contradictory results. Hashizume and Chaussidon (2005) found an $^{16}O$-rich energetic solar wind component, whereas Ireland et al. (2006) found an $^{16}O$ depleted component. The Genesis mission brought back to the Earth specially designed materials implanted with solar wind ions. Despite its catastrophic landing, successful O isotope measurements were performed by McKeegan et al. (2008), which showed an enrichment in $^{16}O$ relative to the Earth of about 7%, slightly larger than that observed in CAIs and comparable within error with the largest $^{16}O$-excess found in a chondrule (~8%, Kobayashi et al. 2003).

### *3.4. Conclusion*

At present, the O isotopic anomaly observed in meteorites seems best explained by self-shielding models (Fig. 9). However, these models are not free from imperfections and the non-mass dependant isotopic fractionation hypotheses are not ruled out. A recent laboratory photochemical experiment resulting in 800% and 400% enrichments in $^{17}O$ and $^{18}O$, respectively, during CO photodissociation without self-shielding (Chakraborty et al. 2008) indicates that the isotopic fractionation associated with CO photodissociation is still far from being understood. Critical isotopic determinations to test these models would be the isotopic composition of CO, which is not available anymore in the solar system, and of water in outer solar system objects, that is in giant planets and cometary ices. In that respect, the determination of oxygen isotopes in water ice from comet Churyumov-Gerasimenko by the Rosetta spacecraft will be of utmost importance. The correct model will have to account for the critical observation that terrestrial planets formed in a major gas reservoir with no isotopic deviation from the Earth, incorrectly attributed to the solar gas by Ozima et al. (2007) but now well characterized in chondrules (Chaussidon et al. 2008) and referred here to as the planetary gas.

## 4. Cold chemistry in the outer nebula : hydrogen isotopes and organosynthesis

### *4.1. Hydrogen isotope systematics in primitive meteorites*

The bulk H isotopic composition of carbonaceous chondrites has long been known to be close to that of the Earth oceans (e.g. Boato 1954, Kolodny et al. 1980, Kerridge 1985, Eiler and Kitchen. 2004). However, following the discovery of large D excesses in unequilibrated ordinary chondrites (Robert et al. 1979, McNaughton et al. 1981), H isotopes were systematically studied in both hydrated minerals and organic matter in ordinary chondrites (e.g. Yang and Epstein 1983, Robert et al. 1987, Deloule and Robert 1995, Deloule et al. 1998) and carbonaceous chondrites (e.g. Robert and Epstein 1982, Kerridge 1983, Kerridge et al. 1987, Halbout et al. 1990, Busemann et al. 2006, Remusat et al. 2006), as well as in interplanetary dust particles (e.g. Zinner et al. 1983, McKeegan et al. 1985, Messenger 2000, Aléon et al. 2001).

These studies have shown a complex hydrogen isotope distribution in unequilibrated ordinary chondrites (UOC) with (1) D-rich water associated with phyllosilicates in the matrix (Deloule and Robert 1995), hydroxylated domains in pyroxene and mesostasis of chondrules (Deloule et al. 1998), (2) D-poor water also found in chondrules (Deloule et al. 1998) and (3) D-rich organic matter (Yang and Epstein 1983, Robert et al. 1987).

In carbonaceous chondrites and interplanetary dust particles, however, organic matter was found to be systematically enriched in D relative to phyllosilicates by at least a factor of two (Fig 10). Isotopic mapping of chondritic interplanetary dust particles (IDPs) has revealed that large D excesses were similarly associated to organic matter while phyllosilicates had lower D/H ratios (Aléon et al. 2001). Isotopic mapping of acid insoluble organic matter have shown that large D excesses are localized into small submicron regions referred to as “deuterium hotspots” (Busemann et al. 2006, Robert et al. 2006).

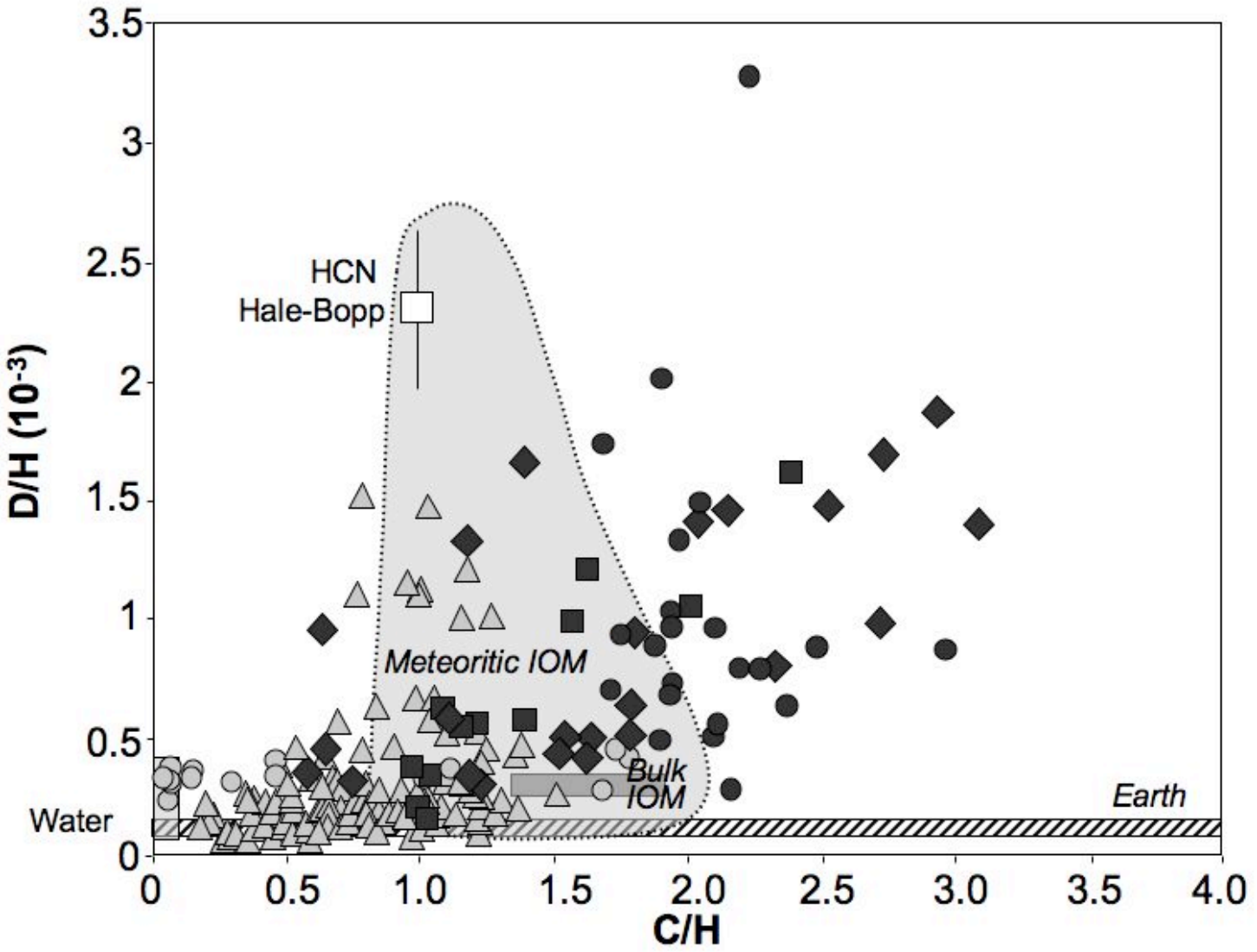

**Fig. 10**. Compilation of D/H and C/H ratios measured in IDPs and hydrated carbonaceous chondrite IOM (CI, CM, CR, hereafter CC) by high spatial resolution isotopic imaging. The distribution indicates mixing between D-poor water of close to terrestrial composition and a complex mix of organic matter enriched in D to variable extent. The most D-rich material in chondrites is comparable to cometary HCN (Meier et al. 1998) and probably contains a significant fraction of D-rich radicals (Gourier et al. 2008). Some IDPs have a composition similar to CC (light grey symbols) but many IDPs, all of which are cluster anhydrous IDPs, contain D-rich components of higher C/H ratios not represented in hydrated carbonaceous chondrites (dark grey symbols). The field of meteoritic IOM is from Busemann et al. (2006) and Robert et al. (2006). IDP data are from Aléon et al. (2001), Mukhopadyay et al. (2003) and Nittler et al. (2006).

### *4.2. Hydrogen isotopes in the galaxy*

The usual explanation advocated for the presence of deuterium-rich organic matter in chondrites is the survival of interstellar molecules from the parent molecular cloud. Despite an initial proposition that large D-excesses could be produced by irradiation (Robert et al. 1979), it was rapidly admitted that the physical conditions in the early solar system preclude large isotopic fractionations (Geiss and Reeves 1981).

Hydrogen isotopes have been extensively measured in the galaxy because both isotopes are produced by Standard Big Bang Nucleosynthesis and the determination of their initial relative abundances is critical to unravel the baryonic density of the universe (e.g. Geiss 1993, Burles 2002). Their abundance in the galactic environments is mainly controlled by two processes : stellar evolution and interstellar chemistry in cold dense clouds.

During stellar evolution, deuterium is converted into $^3He$ by addition of a proton. D-burning is the first nuclear reaction that proceeds when a star ignites and is completed before the star reaches the main sequence, when H-burning starts. D-burning is such a low temperature nuclear process that it can be used to define the minimum mass of starlike objects (e.g. brown dwarfs) even if their mass is not sufficient to start H-burning (e.g. Chabrier et al. 2000). As a result, stars are devoid of deuterium and release D-free matter to the interstellar medium when they die or through winds during their lifetime. The result of this process is that the average D/H ratio of the interstellar medium (ISM) decreases constantly with time. For instance the present-day D/H ratio of the local ISM $H_2$ is estimated to be $1.4 \pm 0.24 \times 10^{-5}$ (Vidal-Madjar 2002), whereas the D/H ratio of $H_2$ ~4.57 Gyr ago, when the solar system formed, is estimated to be $\sim 2.5 \times 10^{-5}$ from the measurement of $^3He$ in the Sun ($2.1 \pm 0.5 \times 10^{-5}$ , Geiss and Gloeckler 1998) and from the atmosphere of Jupiter ($2.6 \pm 0.7 \times 10^{-5}$ , Mahaffy et al. 1998).

By contrast, enormous enrichments in deuterium are observed in molecules produced by ion molecule and grain surface chemistry in cold molecular clouds. At temperatures below 10 K, the isotopic exchange between $H_2$ and ionized molecules is controlled by the difference in zero-point energy between the deuterated species and the hydrogenated species. As a result, isotope exchange reactions such as between HD and $H_3^+$ (reaction 5) are strongly exothermic and result in production of deuterated molecules.

(5) $HD + H_3+ \rightarrow H_2 + H_2D^+ + \Delta E$
with $\Delta E/k = 240 \pm 60$ K (Snell and Wootten, 1979)

These reactions result in the selective pumping of D from the $H_2$ reservoir, which can be considered as infinite, and produce enrichments in D of up to a factor $10^4$. The detection of multiply deuterated species around protostars (e.g. Loinard et al. 2001, Van der Tak et al. 2002, Lis et al. 2002) shows that D-enrichment upon chemistry in cold dense regions is extremely efficient.

The canonical interpretation for the presence of D-rich organic matter in meteorites is the selective preservation of interstellar molecules, which survived solar system formation, planetesimal accretion and geological processes on parent-bodies (e.g. Yang and Epstein 1983, Kerridge 1983, Messenger 2000, Busemann et al. 2006). Isotopic exchange and re-processing of organic matter with water of lower D/H ratio would result in a reduction of the initial ratios from typical interstellar values down to the level at which they are observed. For instance, water at isotopic equilibrium with nebular $H_2$ would have a D/H ratio $\sim 8 \times 10^{-5}$. Such a water may have been preserved in hydroxylated domains in chondrules from ordinary chondrites (Deloule et al. 1998).

### *4.3. Organic matter in carbonaceous chondrites*

Understanding the origin of H isotope variations in the solar system is critically linked to the knowledge of the structure of the organic matter and to the conditions of its synthesis. For obvious exobiological reasons, organic matter has been studied extensively in meteorites and especially in carbonaceous chondrites since the very fall of Orgueil (Berthelot 1868). Only a brief description and the most recent results having implications for H isotopes are given here.

Organic molecules in carbonaceous chondrites can be divided roughly into two groups: simple molecules that are soluble in solvents and acids and insoluble complex macromolecules.

The latter have often been compared with terrestrial kerogens, a term of improper use in this case, and is hereafter referred to as Insoluble Organic Matter (IOM).

Soluble molecules are numerous and include molecules from many chemical families: for instance aliphatic hydrocarbons, polycyclic aromatic hydrocarbons (PAHs), carboxylic and dicarboxylic acids, amines, and biologically relevant molecules such as amino-acids, sugars (polyols) or nucleobases (purines and pyrimidines). In many cases, they host variable D-excesses comparable to or lower than those reported in IOM (e.g. Pizarello et al. 2006). Small enantiomeric excesses have also been reported in amino-acids (e.g. Engel and Macko 1997). Despite their exobiological interest, soluble organic molecules are minor carbon compounds in meteorites and together represent no more than a few 1000 ppm of the total carbon. Their mechanisms of synthesis and their relationship with the IOM are still poorly understood.

IOM represents about 70 to 90 wt% of the total carbon in carbonaceous chondrites, that is up to 3-5 wt% of their host meteorite in CI and CM chondrites. IOM is made of complex macromolecules with thousands of carbon atoms. It is heavily heterosubstituted with several % of N, O and S. Raman spectroscopy and electron microscopy suggest that the structure of IOM becomes more organized as the petrologic type of meteorites increases, indicating an effect from thermal metamorphism in the parent-body (Quirico et al. 2003, Rouzaud et al. 2005, Bonal et al. 2007). The most primitive IOM is found in CI, CM and CR chondrites and has been extensively characterized in Orgueil and Murchison.

The basic structure of IOM is constituted of numerous small aromatic units of 3-4 rings in diameter linked together by small highly ramified aliphatic chains (Gardinier et al. 2000, Cody et al. 2002). The infra-red spectroscopy of the aliphatics in the 3.4 μm region has shown a statistical distribution of $CH_3$ and $CH_2$ groups resulting in a $CH_3/CH_2$ ratio close to 2 (Ehrenfreund et al. 1991). This characteristic is ubiquitous in interstellar organic matter and has been observed along many lines of sight in the galaxy (e.g. Pendelton et al. 1994). By contrast, aromatic units are significantly smaller than those observed in interstellar PAHs, which suggests that small aromatic units in meteoritic IOM escaped photodestruction by UV photons, contrary to interstellar PAHs (Derenne et al. 2005). An important feature of IOM is the presence of organic radicals, i.e. carbon atoms with unpaired electrons, heterogeneously distributed in the IOM (Binet et al. 2002). Mono-radicaloids are known in terrestrial kerogens, but IOM contains also diradicaloids that are unique to extraterrestrial organic matter (Binet et al. 2003). To explain these properties, Derenne et al. (2005) proposed that meteoritic IOM did not originate in the cold ISM but rather formed in an ionized medium at the surface of the protosolar accretion disk, in a layer where the density is high enough to render condensation faster than photodestruction but still low enough to be optically thin.

### *4.4. Nebular origin of the D excesses in organic matter*

The average D/H ratio of the various organic moieties was determined in the IOM using a combination of organic chemistry techniques and mass spectrometry (Remusat et al. 2006). Including the D/H ratio of water, the D/H ratio was found to increase with the following sequence : OH < H bonded with aromatic carbons < H bonded with aliphatic carbons < H bonded with benzylic carbons. This sequence corresponds to a sequence of decreasing bonding energy between H and C atoms. This implies that D is preferentially associated with highly exchangeable hydrogen. Thus it is not stored in a robust molecule that escaped isotopic exchange during solar system and planet formation. Rather, the data are best explained by a late isotopic exchange between a D-poor organic matter/water and a D-rich reservoir, with the most weakly bonded hydrogen having undergone a larger isotopic exchange. In this scenario, the IOM was not significantly reprocessed after its formation. The recent discovery of extreme D excesses in radicals (Gourier et al. 2008) with D/H ratio up to $1.5 \times 10^{-2}$ in benzylic radicals supports this conclusion because the H-C bonds in benzylic radicals have the lowest bonding energy. This D/H ratio typical of ion molecule chemistry suggests that IOM exchanged its H isotopes with an ionized reservoir, possibly $H_2D^+$, formed at very low temperature (< 40 K, Gourier et al. 2008).

Together with the size distribution of aromatic units, this suggests that meteoritic IOM did not form in the parent molecular cloud but rather formed in the solar nebula, possibly at the surface of the disk in a cold ionized medium and was preserved from photodissociation by sinking onto the denser disk midplane (Robert 2002, Derenne et al. 2005). This conclusion is supported by the O isotopic composition in the IOM of Orgueil, which is a typical planetary isotopic composition (Halbout et al. 1990) in agreement with a solar system origin of meteoritic organic matter. However, the "disk-surface" hypothesis remains to be confronted with astronomical observations of $H_2D^+$ at the disk midplane around low mass protostars (Ceccarelli et al. 2004, Ceccarelli and Dominik 2005).

## 5. Irradiation processes recorded in meteorites

### *5.1. General comments on irradiation and meteorites*

The observations summarized in section 2 and 3 indicate that many meteoritic materials underwent irradiation by energetic photons, which drove an active photochemistry. Here, we summarize evidences for nuclear irradiation (e.g. spallation, fusion-evaporation).

There is a vast literature dedicated to the study of meteorite exposure to galactic cosmic rays and solar wind during their residence in an asteroidal/cometary regolith and during their transit in the interplanetary medium (e.g. Eugster et al. 2006). These studies are mostly based on noble gases isotopes and live short-lived nuclei such as $^{10}Be$, $^{26}Al$ or $^{36}Cl$ (e.g. Eugster et al. 2006). In addition to solar system irradiation, chondrules have been shown to contain a significant fraction of stable B isotopes produced by spallation in the ISM before the birth of the solar system (Chaussidon and Robert 1995). Because, both spallation in the ISM and irradiation by recent cosmic rays / solar energetic particles do not address the physical conditions in the early solar nebula, they will not be detailed here.

It has long been suggested that some isotopic anomalies in meteorites may record nuclear irradiation processes in the early solar system. Non thermal nucleosynthetic processes upon MeV proton irradiation were investigated and discarded as the source of the O isotopic anomaly in meteorites (Clayton et al., 1977; Lee 1978). Similarly, the excesses of deuterium discovered in ordinary chondrites were initially attributed to irradiation (Robert et al. 1979) before being attributed to ion-molecule chemistry. Excesses of cosmogenic Ne and Ar relative to those produced by a long term regolith exposure to galactic cosmic rays (GCR) were found in meteoritic minerals having heavy ions irradiation tracks and were first attributed to proton irradiation by a young T-Tauri Sun (Caffee et al. 1987). However, a better estimation of the exposure geometry suggests that there is probably no discrepancy with long term exposure to GCR (Wieler et al. 2000) with the possible exception of minerals from the Murchison meteorite.

### *5.2 Production of extinct short-lived radioactivities by energetic particles from the young Sun*

Short lived nuclei produced long ago are now extinct in meteorites. However, they can still be detected by the analysis of their stable daughter nuclei. Many daughter nuclei have been detected in meteorites, most of which have decay constants long enough to require a production before the birth of the solar system by galactic processes. In only a few cases, local production by irradiation by energetic particles from the protosun have been proposed: $^{7}$Be, $^{10}$Be, $^{26}$Al, $^{36}$Cl, $^{41}$Ca and $^{53}$Mn using particle production deduced from X-ray observations of protostars (e.g. Feigelson et al. 2002).

The discovery of extinct $^{10}$Be in a CAI (McKeegan et al. 2000) provided the first evidence of particle irradiation by the young Sun. Indeed $^{10}$Be has a half-life of 1.5 Myr and is dominantly produced by spallation of $^{16}$O by protons and helium nuclei in the MeV-GeV range. Despite it was proposed that $^{10}$Be could be produced by spallation in the ISM and trapped in the collapsing protosolar cloud core (Desch et al. 2004), the recent discovery in the same CAI of correlated $^{7}$Be excesses also produced by spallation (Chaussidon et al. 2006) provides an unambiguous evidence for a local production because the half-life of $^{7}$Be is only 52 days, which rules out formation before the birth of the solar system. At present, Be isotopes are best explained by irradiation of proto-CAI dust by particles with characteristics of impulsive solar flares rich in $^{3}$He (Gounelle et al. 2006b), with total proton fluences ~$10^{16}$-$10^{18}$ p/cm$^2$ having an energy > 10 MeV (McKeegan et al. 2000). The presence of $^{7}$Be require short irradiation events with calculated fluxes ~$10^{10}$ p/cm$^2$/s with Ep>10 MeV (Gounelle et al. 2006b).

The possible production of other extinct radioactivities and especially of $^{26}$Al by particle irradiation has been investigated in details (e.g. Gounelle et al. 2001, 2006b, Leya et al. 2003) but energetic constraints suggest than only Be isotopes and possibly $^{41}$Ca can be efficiently produced by irradiation (Duprat and Tatischeff 2007). It was recently suggested that injection in the protosolar nebula of $^{26}$Al as well as $^{60}$Fe produced in supernovae may result from fluctuations in the galactic background possibly due to the mechanism of formation of the parent molecular cloud of the Sun (e.g. Gounelle and Meibom 2008, Gounelle et al. 2008) rather than be injected by a supernova shortly before the birth of the solar system.

### *5.3. Stable isotope anomalies produced by nuclear irradiation in the solar nebula?*

Extreme oxygen isotope deviations from the average solar system composition were recently found in micrometer-sized silica-rich grains from the Murchison meteorite (Aléon et al. 2005a), with $^{18}$O/$^{16}$O and $^{17}$O/$^{16}$O ratios reaching 0.1, up to 50 and 200 times larger than usual solar system ratios, respectively. Such large deviations from the solar isotopic composition are usually attributed to stellar nucleosynthesis products preserved in presolar grains (e.g. Clayton and Nittler 2004). However, the isotopic compositions of presolar grains are largely scattered, which is attributed to the multiplicity of their parent stars. By contrast, oxygen isotopes in the exotic silica-rich (ES) grains from Murchison indicate a mixture between a single anomalous source and typical solar oxygen (Aléon et al. 2005a). Similar O isotopic ratios were found in $CO_2$ from the excentric binary post-AGB star HR4049 (Cami et al. 2001), which suggests that such a star may have polluted the young solar system. In this case, O isotopic ratios could be attributed to stellar nucleosynthesis if the companion is a CO white dwarf undergoing nova episodes prior to common envelope evolution (Lugaro et al. 2005). However, the recent re-evaluation of O isotopic composition of HR4049 based on CO analysis suggests that early data were overinterpreted and that HR4049 has O isotopes typical of AGB stars (Hinkle et al. 2007). Furthemore, the absence of Si and Mg isotope anomalies, within 10‰ and 20‰, respectively, in the ES grains (Aléon et al. 2005a,b) rules out nova nucleosynthesis, which is expected to produce

$^{29}$Si and $^{30}$Si excesses together with massive amounts of $^{26}$Al which decays into $^{26}$Mg (e.g. Jose et al. 2004).

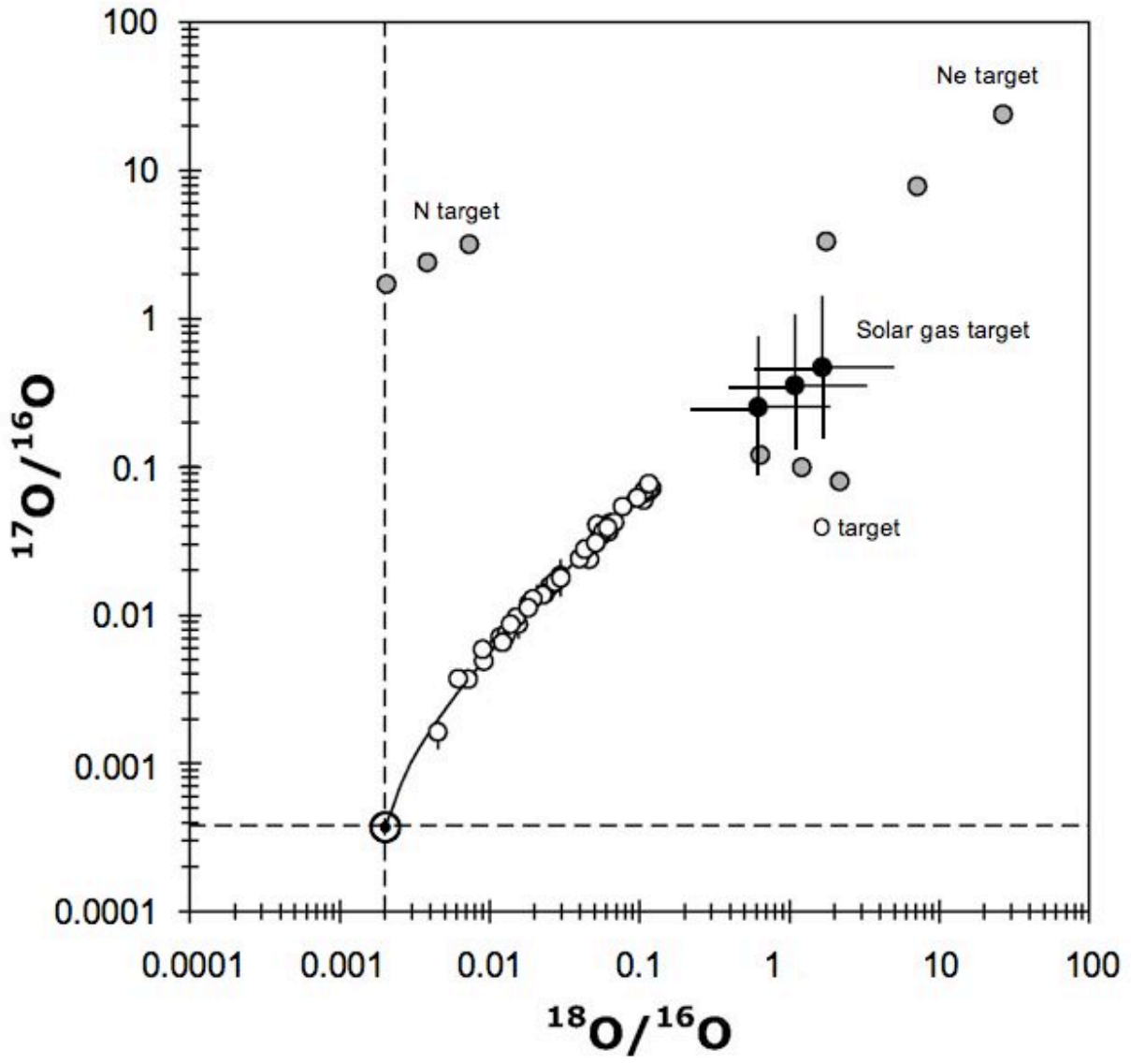

**Fig. 11**. Oxygen isotopic composition of exotic silica-rich (ES) grains together with the oxygen isotopic composition of oxygen produced by irradiation of a N gas, a Ne gas, an O gas and a gas composed of N, Ne and O in solar proportions. Dashed lines show the solar system composition.

The O isotopic composition of the ES grains from Murchison is best explained by non-thermal nucleosynthesis upon irradiation of the circumsolar gas by energetic particles from the young Sun. Excesses of $^{18}$O and $^{17}$O an order of magnitude above those found in the endmember grains are produced by fusion-evaporation reactions at the Coulomb threshold between incident particles with characteristics of $^{3}$He-rich impulsive solar flares and N, O and Ne targets in solar proportions in the gas phase (Aléon et al. 2005a, Fig 11). In order to achieve formation of the ES grains, the nuclear induced oxygen (O*) must be physically isolated from the bulk free O atoms with solar composition and condensation of ES grains must proceed with selective trapping of 0.1 to 10% O*. At present there are no satisfactory explanation for these isolation and selective trapping. An interesting environment of formation to study would be a bipolar jet enriched in SiO by sputtering of silicate dust. In addition, the amount of O* atoms required to account for the concentration of ES grains in Murchison (~1 ppm, Aléon et al. 2005a) is close to the maximum limit possible by irradiation. If the irradiation hypothesis is confirmed, the total proton fluence inferred from the production of $^{18}$O is $10^{16}$-$10^{17}$ p/cm$^2$ with Ep > 10 MeV (Aléon et al. 2005a), which is compatible with that expected from Be isotopes in CAIs (McKeegan et al. 2000, Gounelle et al. 2006b).

### *5.4. Conclusions*

Despite meteorites have a well-known and well-understood record of irradiation by GCR and solar wind during the lifetime of the solar system, and despite repeated propositions of irradiation by protosolar energetic particles, it is only recently that the record of an active T-Tauri phase of the young Sun has been confirmed in meteorites by the discovery of extinct Be isotopes in CAIs (McKeegan et al. 2000, Chaussidon et al. 2006). To what extent protosolar energetic particles could have contributed to other isotopic systems, including stable isotopes, remains to be firmly established. In that respect, understanding the formation of the ES grains of Murchison is essential, because they may provide a record of bipolar outflows from the young Sun (Aléon et al. 2005a).

## 6. Comets in the laboratory : radial mixing and homogeneity in the disk

### *6.1. Cometary rocks in the laboratory*

Cometary samples available in the laboratory can be obtained by two different approaches: (1) terrestrial collection of interplanetary dust particles (IDPs) of possible cometary origin in Earth-crossing orbit and (2) bona fide comet dust recovered in-situ by a spacecraft and returned to the Earth. Both types of samples are nowadays available owing to the Stardust mission, which brought back samples of the short-period comet 81P/Wild2 to the Earth on January $15^{th}$ 2006. Terrestrial collections of IDPs include : (1) Deep sea spherules recovered from the oceanic sediments, (2) IDPs collected by aircraft in the stratosphere, (3) micrometeorites recovered from polar ices and snows and (4) dust particles from space collectors in low altitude orbit. Beside the Stardust samples, the best samples to study the rocky fraction of comets are those which escaped significant alteration on Earth (e.g. thermal metamorphism during atmospheric entry) and in space (hypervelocity impacts in space). They include rare unmelted polar micrometeorites and a significant fraction of stratospheric IDPs. A major issue with terrestrial collections is to discriminate cometary from asteroidal dust particles. In the following, we adopt the traditional designations: dust particles collected in the stratosphere will be referred to as IDPs and dust particles from polar collections will be referred to as micrometeorites and more specifically as antarctic micrometeorites (AMMs). Totally melted dust particles are usually referred to as cosmic spherules (CS), whatever their method of collection.

### *6.2. Chondritic interplanetary dust particles*

Most extraterrestrial IDPs are identified on the basis of their chondritic chemical composition. Because this composition is identical to that of the solar photosphere for rock-forming elements, comet dust is expected to be found among chondritic IDPs. Two major types of IDPs have been found: dominantly hydrated IDPs and dominantly anhydrous IDPs.

#### *6.2.1. Hydrated IDPs: asteroidal?*

Hydrated IDPs are compact “chunky” particles, rather smooth in appearance and are sometimes referred to as “chondritic smooth” IDPs. They are dominated by two types of phyllosilicates: they are either smectite-rich (saponite) or serpentine-rich (e.g. Bradley and Brownlee 1986). They contain abundant FeNi-sulfides (pyrrhotite and pentlandite), carbonates, magnetite of possible framboidal texture and organic matter (4-20 wt%, Keller et al. 1994). Minor minerals include FeNi metal and anhydrous silicates such as olivine, ortho- and clinopyroxene, glass. Hydrated IDPs are similar to but not identical to hydrated carbonaceous chondrites. Only occasionally hydrated IDPs have been identified as CI (Keller et al. 1992) or CM (Bradley and Brownlee 1991, Rietmeijer 1996) type dust. Rare unmelted AMMs have been found to be comparable to hydrated IDPs (Nakamura et al. 2001, Noguchi et al. 2002) and to moderately heated fragments of CI, CM and Tagish Lake carbonaceous chondrites (Nozaki et al. 2006). Their mineralogical similarity with hydrated carbonaceous chondrites and their reflectance spectra similar to CI, CM chondrites and type C asteroids (Bradley et al. 1996) suggests that hydrated IDPs probably sample a set of hydrated carbonaceous chondrite asteroids more representative of asteroid diversity that those available in meteorite collections. We note here that a cometary orbit has been proposed for the CI chondrite Orgueil (Gounelle et al. 2006a) and that Tagish Lake has a reflectance spectrum similar to outer belt type D asteroids (Hiroi et al. 2001), as found for some comets (Licandro et al. 2003, Abell et al. 2005). Most AMMs resemble partially melted hydrated IDPs/carbonaceous chondrites (e.g. Kurat et al. 1994, Engrand and Maurette 1998).

### *6.2.2. Anhydrous IDPs: cometary?*

Anhydrous IDPs are porous, fragile dust particles with aggregate textures of fine-grained sub-micrometer grains and coarser micrometer sized minerals (e.g. Bradley and Brownlee 1986). They are commonly referred to as “chondritic porous” anhydrous IDPs (CPA). They are dominated by forsteritic olivine, enstatitic pyroxene or amorphous silicates, which led to further subdivisions based on infra-red spectra into olivine-rich, pyroxene-rich and glass-rich IDPs (Sandford and Walker 1985), although there is no clear gap between these groups. Other major components include Fe-sulfide minerals (troilite and pyrrhotite) and a cement of organic matter (up to 45 wt%, Keller et al. 1994). Minor constituents are metal, glass, magnetite and refractory minerals. Despite being of chondritic composition, these IDPs have no mineralogical equivalent among anhydrous chondrites. They are possibly related to a new class of porous, fragile AMMs recently discovered in the snows of Central Antarctica and currently under study (Duprat et al. 2007, Engrand et al. 2007, Dobrica et al. 2008). Chondritic IDPs are sometimes found fragmented into hundreds to thousands smaller particles on stratospheric collectors, the so-called cluster IDPs. Despite both hydrated and anhydrous IDPs can be found as clusters, most cluster IDPs are highly porous anhydrous IDPs (e.g. Thomas et al. 1995).

Most amorphous silicates in anhydrous IDPs are rounded grains aggregated to each other, typically around 100 nm in size although it has recently been proposed that they may be aggregates of smaller sub-units (Keller and Messenger 2005). Their composition is close to chondritic at the 100 nm scale and they contain nanometer sized inclusions of metal (more or less oxidized) and sulfides embedded into a glassy silicate matrix of composition intermediate between Mg-olivine and Mg-pyroxene and are thus referred to as Glass with Embedded Metal and Sulfides (GEMS, Bradley 1994). GEMS contain a significant amount of carbon (Brownlee et al. 2000), the most carbon-rich were previously referred to as “tar balls” (Bradley 1988). Some GEMS exhibit a stoechiometric excess of oxygen increasing towards the periphery (Bradley 1994), although this does not seem to be systematic. Relict forsterite and sulfide grains have been found within GEMS (e.g. Bradley et al. 1999). GEMS with relict crystals may have shapes reminiscent of initially larger crystals rendered partially amorphous (Bradley and Dai 2004). The latter properties and the comparison with irradiated forsterite and sulfide led to the proposition that GEMS were interstellar silicates which underwent prolonged irradiation ($\sim 10^8$ years) in the ISM by protons with keV energies typical of interstellar shocks (Bradley 1994). The infra-red signature of GEMS aggregates in the 10 μm regions is indeed comparable with the typical amorphous silicate feature observed in different interstellar environments (Bradley et al. 1999). However, the widely accepted idea that GEMS are interstellar silicates has recently been challenged by sub-micrometer isotopic imaging of GEMS, which indicates that less than 5% of GEMS preserved an isotopic anomaly indicating an origin in RGB/AGB stars (Stadermann and Bradley 2003, Keller and Messenger 2005). Most GEMS have isotopic composition within 10 % of that of the average solar system (Keller and Messenger 2005). Only a small fraction of GEMS would thus have an interstellar origin, while most GEMS would be non-equilibrium condensates from a gas of solar composition (Keller and Messenger 2005). It was also suggested that GEMS could result from partial annealing of totally amorphous interstellar silicates in reducing conditions resulting in nucleation of Fe metal from the reduction of FeO in the silicate glass (Davoisne et al. 2006).

High temperature vapour phase condensates have also been found in IDPs. They include enstatite pyroxene with platelet and ribbon ("whiskers") morphologies (Bradley et al., 1983). These morphologies are typical from an anisotropic growth from a gas phase. The detailed nanoscale crystallography of enstatite whiskers indicate vapour-phase growth guided by crystal defects at T>1300 K (Bradley et al., 1983). Other high temperature condensates include micrometer-sized isolated euhedral forsteritic olivine crystals, which have been reproduced in condensation experiments (Toppani et al. 2006a). Some of these crystals additionally have high Mn content, which is usually attributed to the equilibrium condensation of Mn from the gas into a hot olivine crystal (Klöck et al. 1989). Again estimated condensation temperatures are higher than 1300 K. Finally micrometer-sized CAI minerals have been found in anhydrous IDPs (Christoffersen and Buseck, 1986), some IDPs being exclusively composed of CAI minerals (Zolensky 1987) having typical CAI oxygen isotopic composition (McKeegan 1987).

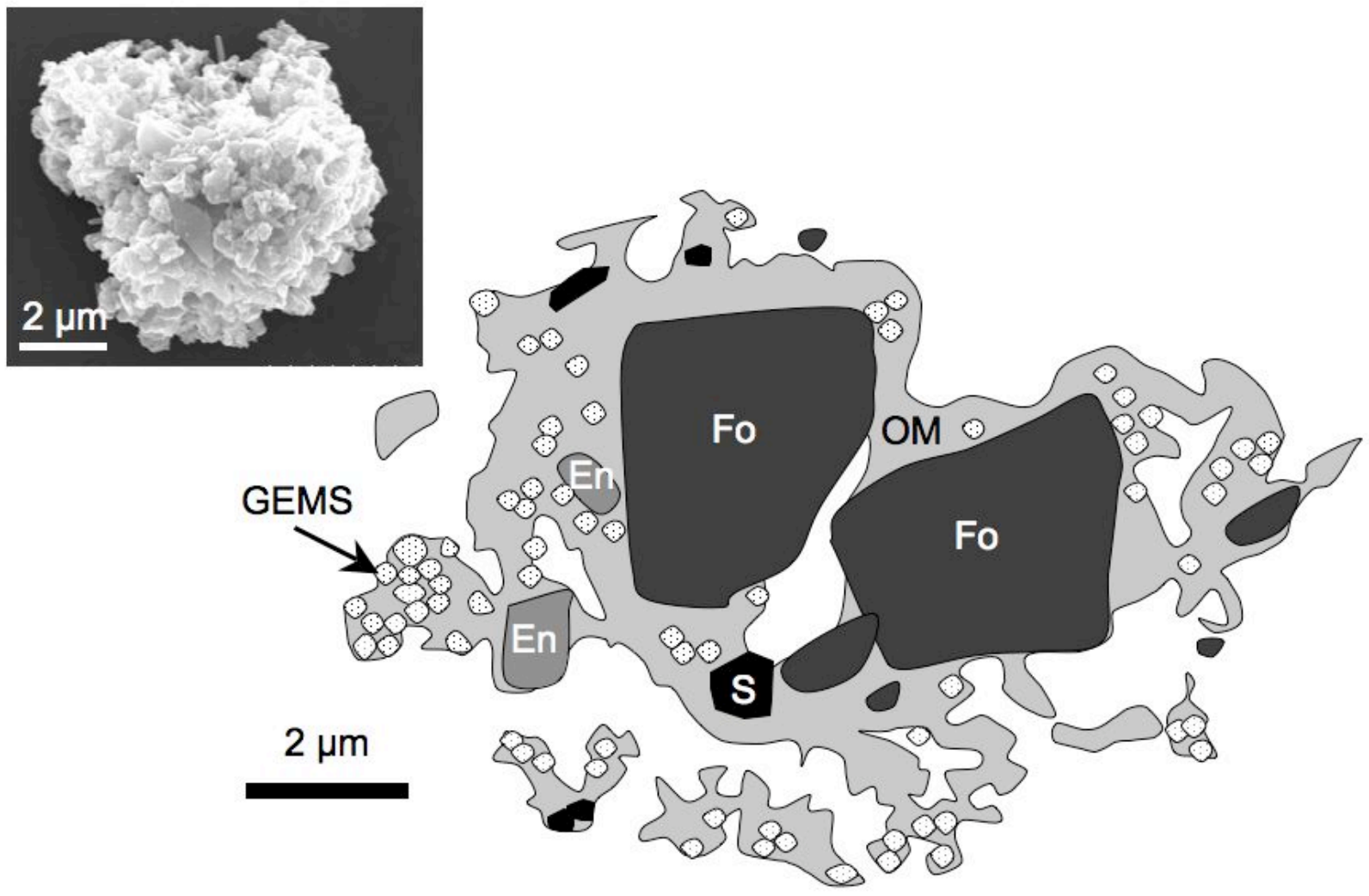

**Fig. 12**. Structure of an ideal cometary dust grain as expected from the study of anhydrous IDPs. Schematics drawn from the transmission electron micrograph of IDP L2009J4 courtesy provided by L.P. Keller (personal communication). OM stands for organic matter, Fo for forsterite, En for enstatite and S for Fe-sulfide. Upper left, a high resolution scanning electron micrograph of cluster IDP L2036W5 shows the typical external morphology of porous anhydrous IDPs.

Anhydrous chondritic IDPs can thus be viewed as highly primitive aggregates of dust grains formed by irradiation, condensation or annealing in protosolar or interstellar environments. In that respect, they tend to have larger amount of D-rich organic matter (e.g. Messenger 2000) and a D-rich organic component that is not observed in meteorites (Fig. 10). They probably contain larger amounts of presolar silicates with RGB/AGB star oxygen isotopic composition (Messenger et al. 2003). A presolar silicate (olivine) of supernova origin has even been found in such an IDP (Messenger et al. 2005).

To summarize anhydrous IDPs have many properties that suggest a cometary origin (Fig. 12):

(1) they are dominated by submicron grains
(2) they are highly porous
(3) they contain up to 45 wt% carbon
(4) they are highly unequilibrated with Fe/(Fe+Mg) ratios in silicates more comparable to those determined from comet Halley than to those of hydrated carbonaceous chondrites (Bradley 1988)
(5) they have infra-red silicate bands in the 10 µm region comparable with various comets (Halley and Bradfield, Bradley et al. 1992, Hale-Bopp, Wooden et al. 2000), which reflects similar relative abundances of amorphous silicates and crystalline Mg-rich olivine and pyroxene (Crovisier et al. 1997, Wooden et al. 1999)

We note in addition that they have reflectance spectra comparable with D-type asteroids (Bradley et al. 1996), as do some Jupiter Family comets (Licandro et al. 2003, Abell et al. 2005)

## *6.3. Comet 81P/Wild2 : Stardust samples*

### *6.3.1. General mineralogy and chemistry*

On January 2nd 2004, the Stardust spacecraft encountered the Jupiter Family comet 81P/Wild2. The encounter occurred at 234 km of the 4.5 km comet nucleus, at 1.86 AU. During the flyby, swarms of dust particles ejected from the nucleus were collected by hypervelocity impact at 6 km/s into a collector composed of 130 silica aerogel cubes of density ranging from <0.01 g/cm$^3$ to 0.05 g/cm$^3$ to achieve a deceleration as gentle as possible. After sample return, examination of the collector indicates that ~10,000 particles of 1-300 µm pre-impact size were successfully collected (Brownlee et al. 2006).

Most samples are recovered from tracks in aerogel showing a large bulb followed by a narrow, deep trail. Comparison with capture experiments (Hörz et al. 2006) suggest that the bulbs were produced by "explosion" of highly porous / highly volatile-rich fine-grained aggregates, whereas the narrow tracks are created and terminated by coarser-grained particles. The spectroscopic analysis of the bulk chemical composition of the bulbs indicates that they are loaded with material of roughly chondritic composition (Flynn et al. 2006). Terminal particles are dominated by (1) coarse-grained (>1 µm) crystalline particles including forsteritic olivine, low-Ca Mg-rich pyroxene (enstatite), FeNi sulfides and to a lesser extent Ca-Al-rich minerals, Na-rich silicates, K-rich feldpar, and Ca-rich pyroxene and (2) amorphous material consisting of vesicular glass, resulting from the melting of aerogel with fine-grained material, and GEMS-like grains (Zolensky et al. 2006). GEMS-like grains have enrichments in Si and sulfides nanograins with core-rim structures, which indicate formation by reaction between aerogel and sulfide grains during high velocity impact capture (Ishii et al. 2008). No hydrated silicates or carbonates were conclusively identified, suggesting the precursor material was dominantly anhydrous. Detailed mineral chemistry indicates strong similarities with typical meteoritic materials and IDPs : Fe/(Mg+Fe) ratios in silicates are comparable to those of anhydrous IDPs and comet Halley, crystalline olivine include Cr- and Mn-rich grains thought to be formed by condensation in the nebula, FeNi sulfides have low Ni content and are more akin to troilite and pyrrhotite from anhydrous IDPs than to Ni-rich sulfides from hydrated IDPs and hydrated carbonaceous chondrites (Zolensky et al. 2006). Detailed chemical imaging of the vesicular glass reveals the presence of sub-μm residual regions with chondritic composition for many elements and with exception of Si (Leroux et al. 2008).

As a whole, chemical and mineralogical analysis of the Stardust samples indicates that the rocky fraction of comet Wild2 has many similarities with anhydrous chondritic objects and more specifically with anhydrous chondritic IDPs (Zolensky et al. 2006), although its study is extremely difficult because the samples were severely damaged during impact capture and highly contaminated with aerogel. It has been claimed that the lack of GEMS and possibly of enstatite whiskers indicates more similarities with asteroidal chondrites than with anhydrous IDPs (Ishii et al. 2008), however fragile sub-μm amorphous silicates were probably preferentially destroyed during impact capture and may have been a significant fraction of the initial comet dust, possibly recorded in areas of chondritic chemistry within melted aerogel (Leroux et al. 2008).

### *6.3.2. Inti : A CAI in a comet*

A spectacular result of the Stardust mission is the discovery of several fragments from a CAI, along one of the first tracks studied (Brownlee et al. 2006, Zolensky et al. 2006). This CAI, nicknamed Inti after the Incan god of the Sun, has been extensively studied (Simon et al. 2008) and exhibit major similarities with usual meteoritic CAIs. The largest fragment was initially ~15 μm in size. It has textures reminiscent of fine-grained unmelted CAIs. Most minerals have μm to sub-μm grain size, although the initial grain size is difficult to assess precisely because only shards produced by ultramicrotome sectioning have been analyzed in detail. It is composed of typical CAI minerals: spinel, Al-rich melilite (gehlenite), Ti-rich clinopyroxene, diopsidic clinopyroxene, anorthite and contains nanoinclusions of metal but also of an unusual Ti-V-rich nitride. Mineral chemistry of pyroxene is similar to that in pyroxene from meteoritic CAIs, notably the Ti valency ($Ti^{III}/Ti^{IV}$) is comparable to that in Ti-rich pyroxene from once-melted coarse-grained type B inclusions from Allende, which suggest formation in similar reducing conditions. The presence of Ti-V-nitride is unusual and if formed by condensation, as usually admitted for fine-grained CAI minerals, requires more reducing conditions than those prevailing at equilibrium in a gas of solar C/O ratio. Nevertheless, Ti-nitride have previously been found in meteoritic CAIs (Weisberg et al. 1988, Bischoff et al. 1989, Meibom et al. 2007). Together with its very low abundance (below 50 ppm) and small size (40 nm), its presence in some CAIs suggests that it could have been missed in other previously studied CAIs. The bulk O isotopic composition of Inti is identical to that of unmelted, unaltered CAIs at $\delta^{18}O \approx \delta^{17}O \approx -40‰$ (McKeegan et al. 2006, Simon et al. 2008). The presence of a CAI similar to meteoritic CAIs in a comet implies radial transport from a hot (T>1500 K) region close to the protosun (<0.1 AU) to cold comet forming regions beyond 20 AU. Several models have been proposed to achieve this transport: transport by turbulence in the accretion disk (Bockelee-Morvan et al. 2002, Ciesla 2007) or ejection from the inner regions by a wind and later decoupling from the gas resulting in the accretion disk being peppered with CAIs (Shu et al. 1996). We note that the presence of CAIs in comets has been predicted as a natural outcome of the X-wind model (Shu et al. 1996), which predicted that small CAIs could be thrown outwards to cometary distances.

### *6.3.3. Isotopic composition of Wild2 dust*

Beside the O isotopic composition of Inti, which is typical of meteoritic CAIs (McKeegan et al. 2006, Simon et al. 2008), O isotopes were measured in several coarse-grained terminal particles consisting of forsteritic olivine and enstatitic pyroxene (McKeegan et al. 2006) and more recently in several terminal particles having strong affinities with chondrule fragments or micro-chondrules (Nakamura et al. 2008). As described in section 2, O isotopes are a powerful tracer of the genetic relationships between planetary materials in the solar system. All Wild2 samples analyzed to date have O isotopic composition identical to their counterpart in carbonaceous chondrites (McKeegan et al. 2006, Nakamura et al. 2008). Furthemore, the isolated olivine and pyroxene crystals have an O isotopic composition very close to that of the planetary gas recorded in chondrules (Chaussidon et al. 2008) and that of the Earth (McKeegan et al. 2006). This indicates that the coarse-grained dust fraction of comet Wild2 is genetically linked with carbonaceous chondrite material and has the typical composition of high temperature inner solar system matter.

Sub-µm O isotopic imaging of many different samples reveals that most matter from comet Wild2 has typical solar system composition within a few % (McKeegan et al. 2006). Only three presolar grains with O isotopic composition indicating formation in a RGB/AGB star outflow, have been found to date (McKeegan et al. 2006, Stadermann and Floss 2008), which indicates that comet Wild2 is made of typical solar system matter and contains interstellar dust grains at the few tens of ppm level (17 ppm, Stadermann and Floss 2008), which is comparable to or lower than chondritic IDPs and carbonaceous chondrites.

H, C and N isotopes have been measured in several organic-rich grains and once-melted aerogel-dust mixture (McKeegan et al. 2006) and have been found to be comparable to those in carbonaceous chondrites with N isotopic anomalies comparable to those measured in IOM and in cometary molecules by remote analysis ($\delta^{15}N$ typically up to ~1500‰, McKeegan et al. 2006, Busemann et al. 2006, Robert et al. 2006, Arpigny et al. 2004, Bockelee-Morvan et al. 2008). D excesses were even found to be lower than those typically measured in chondritic IDPs and chondritic IOM (McKeegan et al. 2006) but this preliminary result needs to be taken with caution because the extent of isotopic contamination by abundant pre-flight terrestrial volatiles trapped in aerogel remains to be evaluated.

All isotopic analyses point to a single conclusion: comet Wild2 is not an aggregate of interstellar matter that survived the collapse of the presolar cloud and the formation of planetary materials as widely thought for many years following the Greenberg model (Greenberg 1982, Levasseur-Regourd 2004). Rather, it is a planetesimal formed from a mix of typical solar system matter, which includes high temperature crystalline material formed in the inner regions of the protosolar nebula (pyroxene, olivine, CAI Inti and possible chondrules). The presence of the latter material in a comet of the Jupiter Family thought to be formed in cold regions beyond Neptune, requires large scale radial transport of matter and a significant amount of mixing and homogenization, regardless of the details of the transport mechanism(s) (e.g. turbulent diffusion, winds).

### *6.4. Carbonaceous chondrites, comets and exo-zodiacal clouds*

The discovery of chondrule fragments (or microchondrules) in comet Wild2 (Nakamura et al. 2008) having O isotopic composition typical of chondrules from carbonaceous chondrites is essential to establish the genetic relationship between comets and carbonaceous chondrite asteroids. If this discovery is confirmed, comet Wild2 must be considered by definition to be a member of the carbonaceous chondrite family. In addition, the oxygen isotopic analysis of chondritic IDPs indicates that they are samples of parent-bodies genetically related to carbonaceous chondrites, including those with a likely cometary origin (Aléon et al. 2008). These two important results suggest that carbonaceous chondrites, primitive outer asteroids (e.g. type C, D, P, Trojans, Centaurs) and comets from various dynamical origins are probably members from a single large family of early solar system planetesimals consisting of aggregated primordial solar system dust.

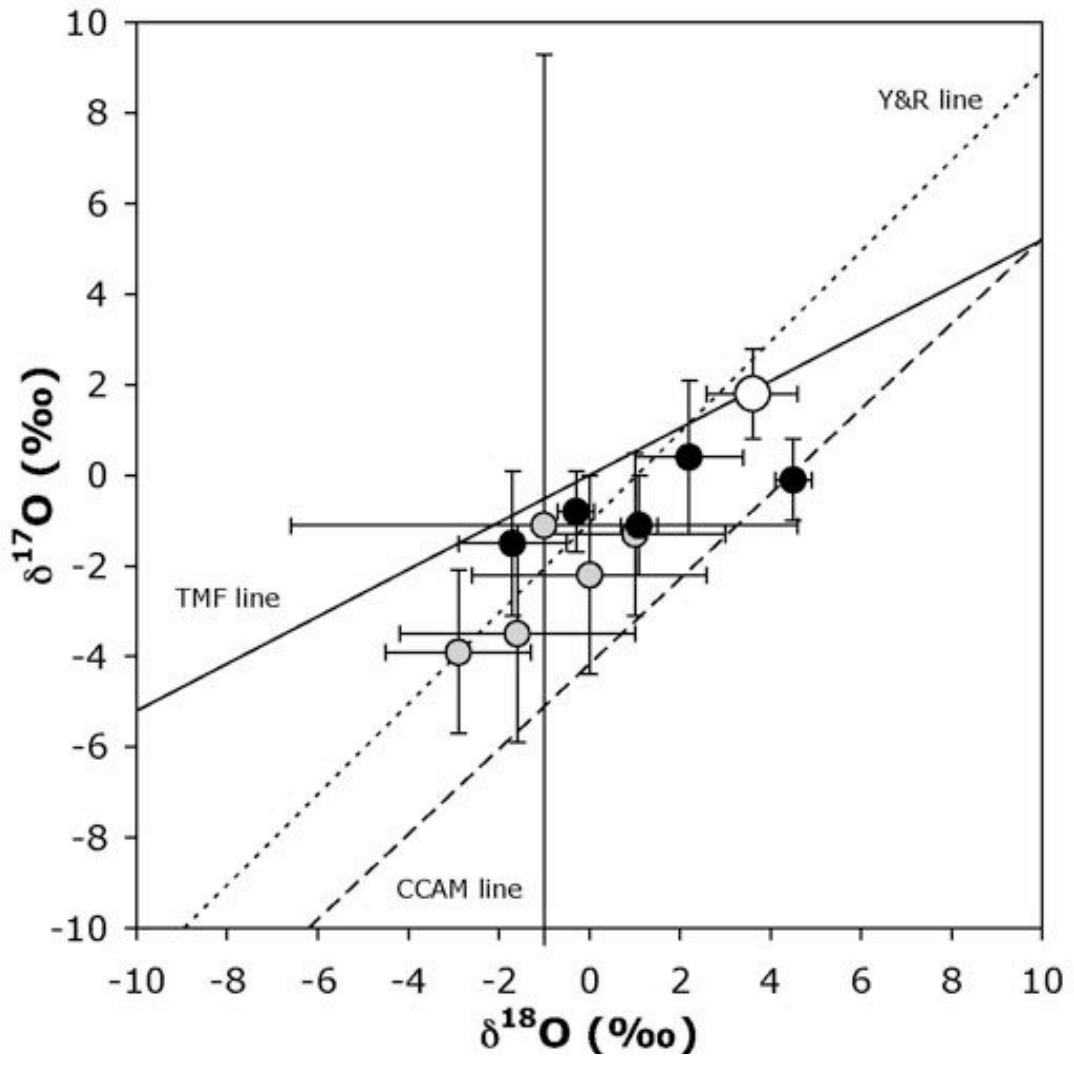

**Fig. 13**. Oxygen isotopes in cometary forsterite and enstatite. Black symbols : Stardust data (McKeegan et al. 2006), grey symbols : anhydrous IDPs data (Aléon et al. 2008), white circle : planetary gas (Chaussidon et al. 2008). TMF stand for Terrestrial Mass Fractionation, CCAM for Carbonaceous Chondrite Anhydrous Minerals and Y&R for the slope 1.0 line defined by Young and Russell (1998).

Dust belts and debris disks have been discovered around other stars (e.g. Smith and Terrile 1984, Kalas et al. 2004), which require dust replenishment by asteroid collisions or cometary activity. The infra-red spectra of these disks as well as those of circumstellar disks around protostars indicates that at least some of them have major similarities with cometary dust from our own solar system: crystalline Mg-rich olivine (forsterite) and pyroxene (enstatite) are detected as well as amorphous silicates, although their respective proportions vary from stars to stars (e.g. Malfait et al. 1998, Kessler-Silacci et al. 2006, Augereau this volume). These proportions are compared with stellar properties to unravel stellar system evolution and the mechanisms of planet formation. In most cases, dust belts and debris disks are located beyond the snow line, which suggests that they are made of materials initially formed early around the protostar, accreted as asteroidal or cometary planetesimals and later released in a similar manner to the solar system zodiacal cloud. Because dust from the solar system zodiacal cloud is available in the laboratory as carbonaceous chondrite/cometary dust, it is certainly the best analog to study the properties of dust in other stellar systems.

For instance, the annealing of GEMS at 1000 K results in formation of equilibrated aggregates of FeMg-bearing olivine and pyroxene embedded in a silica-rich glass (Brownlee et al. 2005). These equilibrated aggregates are also commonly found in anhydrous IDPs. They were thus probably GEMS annealed during atmospheric entry. We note that if GEMS really are interstellar silicates their annealing do not result in individual crystals of forsterite and enstatite such as those found in comets and circumstellar disks, which suggests the latter do not form by annealing of interstellar dust (Bockelee-Morvan et al. 2002). In addition, the oxygen isotopic composition of forsterite and enstatite in anhydrous IDPs and comet Wild2 is comparable to that of chondrules and is close to that of the gas in which chondrules were formed (Aléon et al. 2008, Fig 13). Some of these crystals may be considered as chondrule fragments or microchondrules (Nakamura et al. 2008, Aléon et al. 2008). However, many individual forsterite and enstatite minerals in anhydrous IDPs have euhedral and whisker crystal shapes, which suggest that they form at high temperature by condensation. Mineralogical and isotopic properties of crystalline forsterite and enstatite in IDPs thus suggest that their counterpart in circumstellar disks formed in the inner regions close to the protostars, rather than be annealed interstellar grains.

## 7. Recent developments in chronology

The chronology of solar system formation has been extensively studied using the isotopic composition of several radiogenic elements in meteorites (e.g. McKeegan and Davis 2005). We note that many details are still to be firmly established because of problems in (1) the intercalibration of the various isotope chronometers, (2) the anchoring of the relative chronometers based on extinct radioactivities to the absolute ages given by long-lived radionuclides and (3) uncertainties in the homogeneity of the parent-nuclides distribution in the solar nebula. McKeegan and Davis (1995) give the following chonology:

4.568 Gyrs ago : Beginning of the collapse of the presolar cloud, formation of “anomalous” CAIs before incorporation of radionuclides of stellar origin

4.567 Gyrs ago : CAI formation during a few $10^5$ years

4.566 Gyrs ago : Chondrule formation starts and lasts for 1-2 Myrs

4.565 – 4.564 Gyrs ago : Accretion of chondritic planetesimals, large scale melting in the parent body of differentiated HED meteorites (asteroid 4Vesta).

Igneous activity and metamorphism on the parent-bodies of differentiated metorites lasted for a few tens of million years starting from 4.564 Gyrs ago.

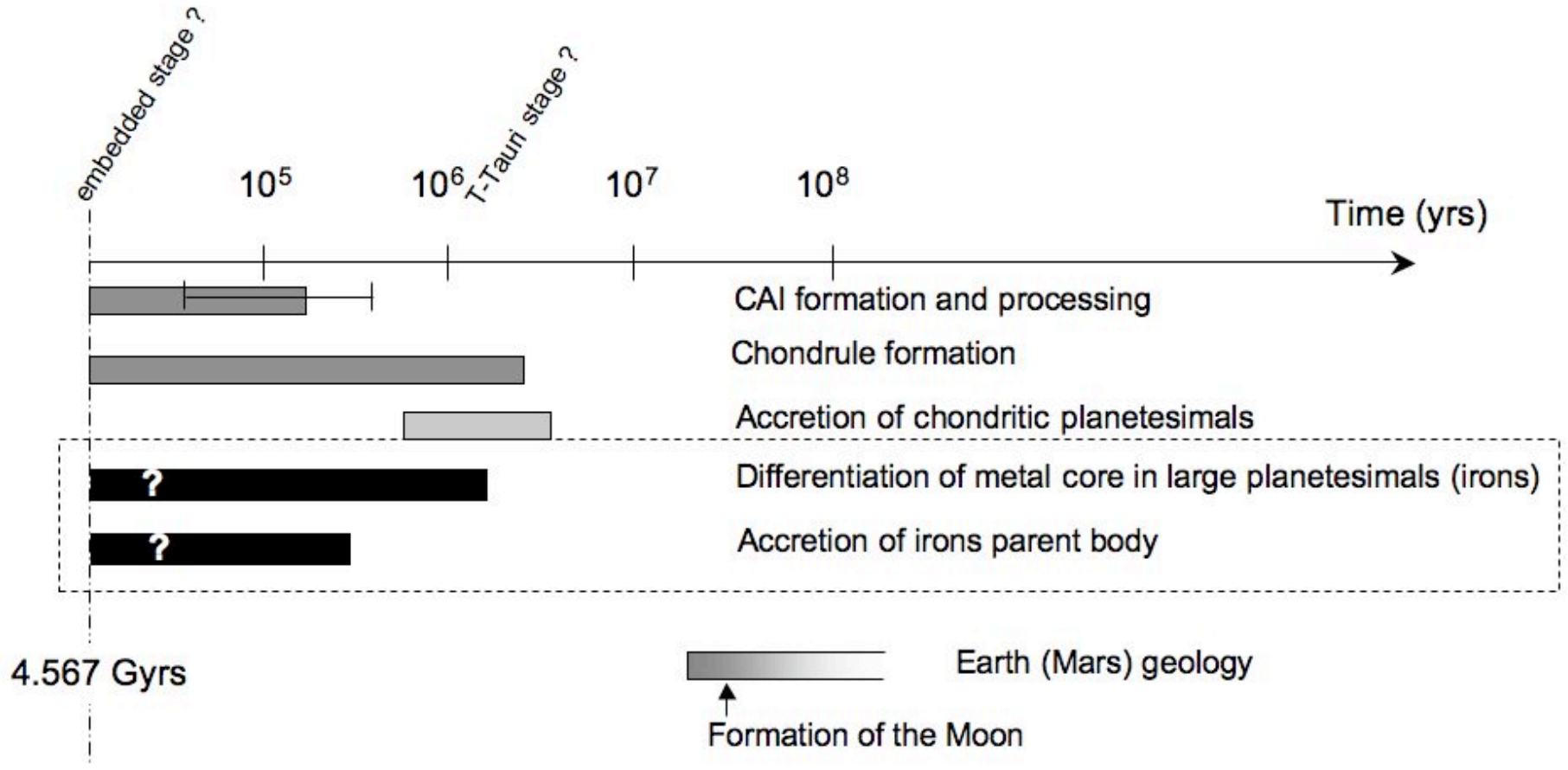

**Fig. 14.** Chronology of solar system formation

Recent modifications and additions to this scenario are mentioned here (Fig. 14). Bulk high precision Al/Mg dating of a large number of CAIs suggest that $^{26}Al$ was homogeneously distributed in the early solar system and confirm the validity of Mg isotopes as a chronometer (Thrane et al. 2006, Connelly et al. 2008). Combined with high precision analyses of Pb isotopes, they establish a relative chronology of CAI and chondrule formation (Connelly et al. 2008). Bulk isotope analysis suggests that the duration of CAI formation could be as short as 20,000 years and probably no more than 100,000 years (Thrane et al. 2006). However, localized isotopic analyses indicate an extended period of CAI crystallization, up to several $10^5$ years (e.g. Young et al. 2005). It is thus currently believed that CAIs formed extremely early during solar system formation, possibly during an embedded stage and possibly during the initial phase of cloud collapse (Thrane et al. 2006) but experienced brief high temperature events during a longer period of time (Young et al. 2005). Chondrule formation is now thought to have started as early as CAI formation (Bizarro et al. 2004) and lasted for about 3 Myr (Connelly et al. 2008).

High precision Pb isotope age of 4.566 Gyrs of the differentiated meteorites angrites (Baker et al. 2005) and excesses of radiogenic $^{26}$Mg due to the decay of $^{26}$Al isotope ratios in HED meteorites (Bizarro et al. 2005) indicate that achondrites (i.e. meteorites from differentiated planetesimals) could have accreted and melted 1 to 3 Myrs after CAI formation during the period of chondrule formation. The early incorporation of large amount of live $^{26}$Al and $^{60}$Fe provided a large amount of heat for a rapid differentiation. Recent $^{53}$Mn/$^{53}$Cr dating of bulk carbonaceous chondrites gives an age comparable to that of CAIs (Moynier et al. 2007) indicating that the accretion of chondritic planetesimals could have been completed shortly after CAI formation. We note that such a rapid accretion of carbonaceous chondrite parent-bodies is in contradiction with an extended period of time for chondrule formation. Furthermore, recent Hf-W ages of magmatic iron meteorites thought to come from cores of proto-planetesimals disrupted by giant impacts indicate that these magmatic iron meteorites crystallized about 1 Myr after CAI formation, which requires accretion during CAI formation and differentiation of planetesimals large enough to undergo core formation during the first million year of the solar system (Kleine et al. 2005, Schersten et al. 2006). Together with the detection of exoplanets still embedded in a protoplanetary disk (Setiawan et al. 2008), and despite some level of contradiction between the various chronological tools, all these results points to an extremely rapid planet forming process, not only for gaseous giant planets, but also for rocky terrestrial planets and suggest that planet formation was contemporaneous to the accretion disk lifetime.

## 8. Conclusion

As exemplified here, the detailed study of chondritic meteorites and cometary dust can bring precise physico-chemical constraints such as pressure, temperature, opacity, photons or particle fluxes, concentrations etc., as well as precise time constraints, to understand the birth and evolution of the young solar system and the formation of planetesimals and planets. However, these informations are highly localized in space and time and a general understanding of the early solar system physical processes requires to put meteoritic studies into a global astrophysical context. This context can be provided by astronomical observations of low-mass protostars and exoplanets or astrophysical models of protoplanetary disks around low mass stars. In return, meteoritic observations can help placing constraints on astrophysical mechanisms occurring in protoplanetary disks and thus provide a better understanding of general processes associated with star and planet formation. As a whole, as shown in the cartoons of appendix 1, meteoritic observations satisfactorily fit with physical processes expected in a protoplanetary disk and suggest that the formation of rocky planetesimals was an extremely rapid process. Once in an astrophysical context, they can be used as detailed tools to probe our distant past and the origin of the solar system.

I would like to thank Thierry Montmerle and the organizing committee for inviting me to give this lecture in Les Houches. Numerous stimulating discussions with Alice Aléon-Toppani and Michèle Bourot-Denise were appreciated. Robert N. Clayton, Andrew M. Davis, Sujoy Mukhopadhyay, Larry R. Nittler and Lindsay P. Keller are warmly thanked for providing materials used for preparation of figs. 6 and 7 (RNC & AMD), fig. 10 (SM & LRN) and fig. 12 (LPK). Sections from Efremovka and Murchison used for figs. 2 and 3 are from the Paris Museum National d'Histoire Naturelle. CAI 101,1 used for fig. 3 was provided courtesy by Ahmed El Goresy.

**Appendix 1. Structures and processes in the solar protoplanetary disk deduced from meteoritic observations**

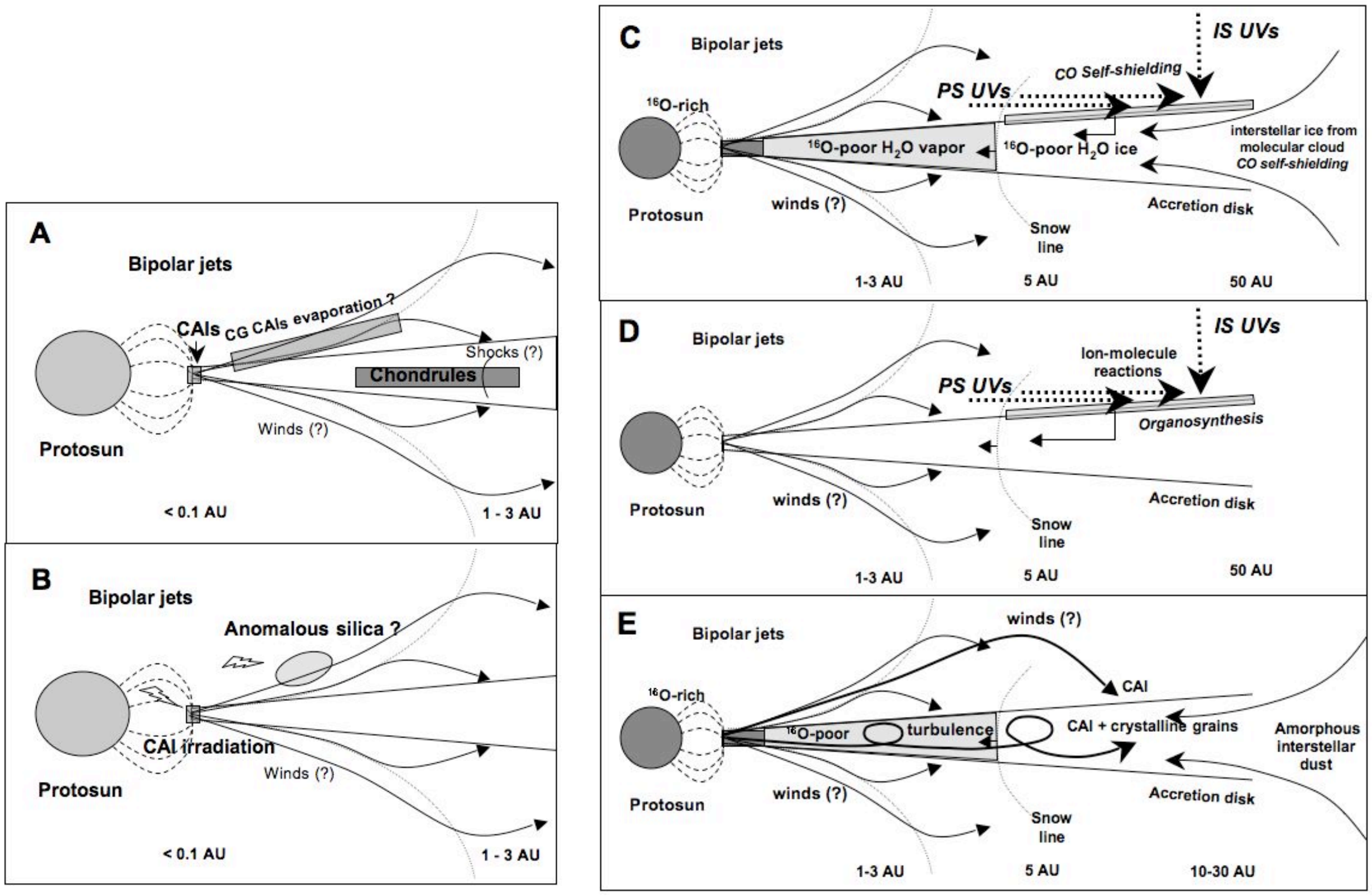

A. High temperature formation of chondrules and CAIs in the inner disk.

B. Irradiation of CAIs/proto-CAIs at the inner edge of the disk, or at the base of the bipolar jet (anomalous silica ?)

C. Structures and transport deduced from O isotopes and self-shielding models. IS UVs : interstellar UV photons, PS UVs : protosolar UV photons

D. Context for organosynthesis and H isotopes fractionation

E. Large scale transport of high temperature minerals found in comet Wild2 and cometary IDPs

Dotted lines : bipolar flows at the origin of wind/disk wind able to transport dust material
Dashed lines : funnel flows
Abbreviations as in main text.

## Appendix 2. Simplified classification of chondrites and other materials relevant to the study of the solar nebula

### Petrologic type

| ← Increasing aqueous alteration | | unaltered | Increasing thermal metamorphism → | | |
|---|---|---|---|---|---|
| **1** | **2** | **3** | **4** | **5** | **6** |

### Chondrite families

| Meteorite groups | Major caracteristics | Famous meteorites[1] | Petrologic type[2] |
|---|---|---|---|
| *Carbonaceous chondrites* | | | |
| CI | hydrated | Orgueil, *Ivuna* | 1 |
| CM | hydrated | Murchison, *Mighei* | (1) 2 |
| CR | hydrated | *Renazzo* | (1) 2 |
| CV | anhydrous | Allende, *Vigarano*, Efremovka, Leoville | 3 (4) |
| CO | anhydrous | *Ornans* | 3 (4) |
| CK | anhydrous | *Karoonda* | (3) 4 - 6 |
| CH | metal-rich | *ALH 85085** | 3 |
| CB | metal-rich | *Bencubbin* | 3 |
| Ungrouped | hydrated | Tagish Lake | 1/2 |
| Ungrouped | anhydrous | Acfer 094 | 3 |
| *Ordinary chondrites* | | | |
| LL | low Fe, low $Fe^0$ | Semarkona, Bishunpur | 3 - 6 |
| L | low Fe | | 3 - 6 |
| H | high Fe | | 3 - 6 |
| *Enstatite chondrites* | | | |
| EL | low Fe | | 3 - 6 |
| EH | high Fe | | 3 - 6 |
| *R-chondrites* | | *Rumuruti* | 3 - 6 |
| *K-chondrites* | | *Kakangari* | 3 |

* ALH stands for Allan Hills in Antarctica, from where the ALH meteorite series are collected
1 Lithotypes from which the groups are named (carbonaceous chondrites and R-K-chondrites) are given in italics, other meteorites are abundantly studied specimen or are cited in the present chapter
2 Parentheses stand for minor amount of meteorites, dashed line give the range and slash indicates controversy in the classification

### Other primitive samples

Interplanetary dust particles – related to carbonaceous chondrites
- Polar micrometeorites – related to CI, CM, CR and Tagish Lake
- Stratospheric IDPs
  - Hydrated – related to CI, CM, CR and Tagish Lake
  - Anhydrous – no exact equivalent in anhydrous carbonaceous chondrites

Comet 81P/Wild 2 (Jupiter Family) – related to anhydrous interplanetary dust particles and potentially to anhydrous carbonaceous chondrites

## Appendix 3. Common minerals in primitive meteorites

| Mineral family | Mineral name[1] | Structural formula | Occurrence[2] |
|---|---|---|---|
| *Silicates* | | | |
| *Anhydrous silicates* | | | |
| Olivine | Forsterite | $Mg_2SiO_4$ | Chd, CAI, Mtx |
| | Fayalite | $Fe_2SiO_4$ | Mtx, solid sol.[3] |
| Pyroxene | Enstatite (opx) | $MgSiO_3$ | Chd |
| | Diopside (cpx) | $CaMgSi_2O_6$ | CAI, Chd |
| | "Fassaite" (cpx) | $Ca(Mg,Ti,Al)(Al,Si)_2O_6$ | CAI |
| Plagioclase feldspar | Anorthite | $CaAl_2Si_2O_8$ | CAI, Chd |
| | Albite | $NaAlSi_3O_8$ | solid sol.[4] |
| Melilite | Gehlenite | $Ca_2Al_2SiO_7$ | CAI |
| | Åkermanite | $Ca_2MgSi_2O_7$ | CAI |
| *Phyllosilicates (hydrous)* | | | |
| Serpentine | Antigorite | $Mg_3Si_2O_5(OH)_4$ | Mtx |
| | Cronstedtite | $Fe^{II}{}_2Fe^{III}{}_2SiO_5(OH)_4$ | Mtx |
| Smectite | Saponite | $(Ca,Na)_{0.3}(Mg,Fe)_3(Si,Al)_4O_{10}\ (OH)_2.4H_2O$ | Mtx |
| *Oxides* | | | |
| | Corundum | $Al_2O_3$ | CAI |
| | Hibonite | $CaAl_{12}O_{19}$ | CAI |
| | Perovskite | $CaTiO_3$ | CAI |
| Spinel | Spinel s.s. | $MgAl_2O_4$ | CAI |
| | Magnetite | $Fe^{II}Fe^{III}{}_2O_4$ | Chd, Mtx |
| | Chromite | $FeCr_2O_4$ | Chd, Mtx |
| *Sulfides* | Troilite | FeS | Chd, Mtx |
| | Pyrrhotite | $Fe_{1-x}S$ | Chd, Mtx |
| | Pentlandite | $(Fe,Ni)_9S_8$ | Mtx |
| *Carbonates* | Calcite | $CaCO_3$ | Mtx |
| | Dolomite | $MgCa(CO_3)_2$ | Mtx |
| *Metal alloys* | Kamacite | $\alpha$-(Fe,Ni) (4-10 wt% Ni) | Chd, Mtx, CAI |
| | Taenite | $\gamma$-(Fe,Ni) (25-40 wt% Ni) | Chd |

(1) abbreviations = opx : orthopyroxene ; cpx : clinopyroxene
(2) abbreviations = CAI : Ca-Al-rich inclusions; Chd : Chondrules; Mtx : Matrix
(3) Fayalite is found isolated in the matrix but it is commonly found in solid solution with forsterite to form Mg,Fe-bearing olivine, abundant in chondrules and the matrix
(4) Albite is barely found isolated but is present in chondrules in solid solution with anorthite to form a calco-sodic plagioclase.